\documentclass[reprint,aps,prb,amsmath,amssymb,twocolumn,showkeys, superscriptaddress,floatfix,showkeys]{revtex4-2}
\usepackage{graphicx}
\usepackage{dcolumn}
\usepackage{bm}
\usepackage[colorlinks = true, allcolors = blue]{hyperref}
\usepackage{verbatim}
\usepackage{soul}
\usepackage{float}
\usepackage{xcolor}
\usepackage{multirow}

\begin{document}

\title{Analytical exciton energies in monolayer transition-metal dichalcogenides}

\author{Hanh T. Dinh}
\affiliation{%
 Computational Physics Key Laboratory K002, Department of Physics, Ho Chi Minh City University of Education, Ho Chi Minh City 72759, Vietnam
}%
\thanks{H.T.Dinh and N.-H.Phan contributed equally to this work.}

\author{Ngoc-Hung Phan}
\affiliation{%
 Computational Physics Key Laboratory K002, Department of Physics, Ho Chi Minh City University of Education, Ho Chi Minh City 72759, Vietnam
}%

\author{Duy-Nhat Ly}
\affiliation{%
 Computational Physics Key Laboratory K002, Department of Physics, Ho Chi Minh City University of Education, Ho Chi Minh City 72759, Vietnam
}%

\author{Dai-Nam Le}
\affiliation{%
 Department of Physics, University of South Florida, Tampa, FL 33620, United States of America
}%

\author{Ngoc-Tram D. Hoang}
\affiliation{%
Computational Physics Key Laboratory K002, Department of Physics, Ho Chi Minh City University of Education, Ho Chi Minh City 72759, Vietnam
}

\author{Nhat-Quang Nguyen}
\affiliation{%
Computational Physics Key Laboratory K002, Department of Physics, Ho Chi Minh City University of Education, Ho Chi Minh City 72759, Vietnam
}%

\author{Phuoc-Thien Doan}
\affiliation{%
Computational Physics Key Laboratory K002, Department of Physics, Ho Chi Minh City University of Education, Ho Chi Minh City 72759, Vietnam
}%

\author{Van-Hoang Le}
\email{hoanglv@hcmue.edu.vn}
\affiliation{%
 Computational Physics Key Laboratory K002, Department of Physics, Ho Chi Minh City University of Education, Ho Chi Minh City 72759, Vietnam
}%

\date{\today}

\begin{abstract}
We have developed an analytical expression for the $s$-state exciton energies in monolayer transition-metal dichalcogenides (TMDCs): $E_{ns}=-{\text{Ry}}^*\times P_n/{(n-0.5+0.479\, r^*_0/\kappa)^2}$, $n=1,2,...$. Here, $r^*_0$ and $\kappa$ represent the dimensionless screening length and dielectric constant of the surrounding medium, respectively. $\text{Ry}^*$ is an effective Rydberg energy scaled by the dielectric constant and exciton reduced mass, and $P_n(r^*_0/\kappa)$ is a function of variables $n$ and $r^*_0/\kappa$. Its values are around 1.0, so it is considered a term that corrects the Rydberg energy. Despite the simple form, the suggested formula provides exciton energies with high precision compared to the exact numerical solutions and accurately describes recent experimental data for a large class of TMDC materials, including WSe$_2$, WS$_2$, MoSe$_2$, MoS$_2$, and MoTe$_2$. To achieve these results, we have developed a regulated perturbation theory by combining the conventional perturbation method with several elements of the Feranchuk-Komarov operator method. This includes the Levi-Civita transformation, the algebraic calculation technique using the annihilation and creation operators, and the introduction of a free parameter to optimize the convergence rate of the perturbation series. This universal form of exciton energies could be beneficial in various physical analyses, including retrieval of the material parameters such as reduced exciton mass and screening length from the available measured exciton energies.
\end{abstract}


\maketitle

\textit{Introduction --} In the past ten years, energy levels of excitons in different monolayer materials have been measured and studied experimentally and  theoretically \cite{berkelbach2013,chernikov2014,Plechinger-2016,Stier2016-nano, Olsen2016, Molas2019, Hieu2022,Liu2019,PRL2018, NAT2019, Chen2019-nano, Wu2019, Arora2021,PhysE, Nhat2022, PhysRevB2023,Sekaria2024,Hansen2024}. Pioneer studies \cite{berkelbach2013, chernikov2014} suggest that, unlike 3D bulk semiconductors, in 2D atomically thin films like monolayer transition-metal dichalcogenides (TMDCs), electrons and holes coupled to each other by what is known as Rytova-Keldysh interaction \cite{Rytova1967, haramura1988,keldysh1979, cudazzo2011} when they form neutral excitons. This interaction demonstrates a short-range logarithmic potential $\ln(r)$, but becomes a conventional Coulomb interaction $1/r$ in the long range \cite{haramura1988, cudazzo2011}. As a result, the energies of low-lying exciton states (where $n < 5$) do not follow the Rydberg series of a two-dimensional hydrogen atom. Therefore, developing an accurate theory of exciton binding energies in 2D semiconductors is essential to benchmark experimental results.

Recently, various methods have been developed to determine exciton binding energies when the reduced mass, screening length, and dielectric constant of the surrounding medium are known \cite{Wu2019, PhysE, Nhat2022, PhysRevB2023}. Our packages \cite{PhysE, Nhat2022, PhysRevB2023}, which are based on the Feranchuk-Komarov operator method \cite{Feranchuk1982, Hoangbook2015}, perform exceptionally well in calculating exciton binding energies in monolayer TMDCs, even in high magnetic fields. On the other hand, theoretical efforts have been made to derive analytical formulas for exciton binding energies in 2D semiconductors \cite{Olsen2016, Molas2019, Hieu2022}. These formulas result from modeling the electron-hole interaction potential in 2D semiconductors to make the associated Wannier Schr\"{o}dinger equation analytically solvable. This approach leads to easily applicable formulas that are generalized from the Rydberg series with reasonable accuracy. However, the accuracy of these formulas is, of course, limited by the accuracy of their modified electron-hole potentials. Therefore, a comprehensive theory is required to analytically describe exciton energies.

In this Letter, we present a highly accurate analytical expression for the energies of $s$-state excitons using a new approach called regulated perturbation theory (RPT). The formula we obtained for exciton energies is simple but universally applicable to a wide range of TMDC monolayers. This formula is precise compared to the exact numerical solutions and describes the experimental data that is currently available. The success of this approach is attributed to its advancements. Unlike conventional perturbation theory, the RPT approach incorporates the Levi-Civita transformation, which enables an algebraic calculation technique and introduces a free parameter to control the convergence rate of the perturbation correction series. Furthermore, it eliminates the need to modify the electron-hole potential and accurately computes all matrix elements associated with the harmonic oscillator basis set. 

\textit{Regulated perturbation theory --} For energy spectra of an exciton in a TMDC monolayer, most studies consider the Schr{\"o}dinger equation
\begin{eqnarray}\label{eq1}
\left\{ -\left(\frac{\partial^2}{\partial x^2}+\frac{\partial^2}{\partial y^2} \right)
     + {\hat V}_{h-e}(r)  -E\right\} \psi(x, y) = 0,
\end{eqnarray}
where $r=\sqrt{x^2+y^2}$; energy $E$ and coordinates $x, y$ are given in the effective Rydberg energy Ry$^{*}=\mu e^4/32\pi^2\varepsilon_0^2\kappa^2\hbar^2$ and Bohr radius $a_0^{*}=4\pi\varepsilon_0\kappa\hbar^2/\mu e^2$ scaled  by the exciton reduced mass $\mu$ and a dielectric constant of the surrounding medium $\kappa$. The electron-hole interaction $\hat V_{e-h}$ is described by the Rytova-Keldysh potential, expressed via Struve and Bessel functions in most studies. Differently, we use its Laplace form as
\begin{eqnarray}\label{eq2}
{\hat V}_{e-h}(r)=-2\int\limits_{0}^{+\infty}dq\frac{e^{-qr}}{\sqrt{1+\alpha^2 q^2}},
\end{eqnarray}
where $\alpha=r_0/\kappa a_0^*$ with $r_0$ - a screening length. 

The work \cite{PhysRevB2023} suggests transforming the Schr{\"o}dinger equation from $(x, y)$ to $(u, v)$ space by the Levi-Civita transformation \cite{giang1993}. The new equation in $(u, v)$ space,
\begin{eqnarray}\label{eq3}
(\hat H- E \hat R)\psi(u, v) = 0,
\end{eqnarray}
 is excluded from the Coulomb singularity, making solving it more convenient. Notably, it allows the solving process to use the harmonic oscillator basis set $|n,m (\omega)\rangle$  with running indices $n=0,1,2,\cdots$ and $m=0, \pm 1, \pm 2, \cdots, \pm n$ (the principal and magnetic quantum numbers). The harmonic frequency $\omega$ can be used as a free parameter to regulate the convergence rate of the perturbation series so that the desired precision is granted. Moreover, the equation in $(u, v)$ space and the harmonic basis set can be presented via the creation and annihilation operators: $\hat a^+, \hat a$ and $\hat b^+, \hat b$. All matrix elements necessary for solving Eq.~\eqref{eq3}, such as  $h_k=\langle n|\hat H|n \rangle$, $r_k=\langle n|\hat R|n \rangle$, $v_{jk}=\langle j|\hat H|k \rangle$, and $v^{\text{\tiny{\it R}}}_{jk}=\langle j|\hat R|k \rangle$ $(j\neq k)$, can then be calculated algebraically using the commutation relations $[\hat a, \hat a^+]=1$ and $[\hat b, \hat b^+]=1$. Here, we consider only $s$-states where $m=0$, so the basis set can be shortly written as $|n\rangle \equiv |n, m=0\rangle$.  This algebraic approach obtained analytical forms of these matrix elements. We provide them in Supplement~\cite{Suppl} for use in the present study. 

Different from Ref.~\cite{PhysRevB2023}, where Eq.~\eqref{eq3} was treated numerically with highly accurate exciton energies, we solve this equation analytically by the perturbation method, expanding wave functions and energies by the perturbation parameter $\beta $ as 
$\psi=\psi^{(0)}+\beta\Delta \psi^{(1)}+\beta^2\Delta \psi^{(2)}+\cdots$, 
$E=E^{(0)}+\beta\Delta E^{(1)}+\beta^2\Delta E^{(2)}+\cdots$. Plugging these expansions into Eq.~\eqref{eq3}, where the operators are separated as $\hat H = \hat H_0 +\beta\, \hat V$ and $\hat R = \hat R_0+\beta \hat V_R$, we obtain formulas for the wave function $\psi^{(0)}$ and energy $E^{(0)}$ in the zero-order approximation and their corrections $\Delta \psi^{(s)}$, $\Delta E^{(s)}$ in any approximation order $s=1,2,3,\cdots$. Energies in the $s$-order approximation are calculated by equation $E^{(s)}= E^{(0)} + \Delta E^{(1)} + \Delta E^{(2)} +\cdots + \Delta E^{(s)}$. The present study is interested in exciton energies only to the third-order approximation, whose formulas are provided in Supplement~\cite{Suppl}.

Table~\ref{tab1} shows exciton energies in monolayer WSe$_2$ in the zero-, second-, and third-order approximations for $1s$ to $5s$ states, calculated by the regulated perturbation theory and compared with the numerically exact solutions \cite{PhysRevB2023}. The free parameter $\omega$ is chosen by the variational condition $\partial E^{(0)}/\partial \omega=0$ for these energies. As shown in Table~\ref{tab1}, the energy precision is very high, within 1.0 meV for the second-order approximation, while less than 0.2 meV for the third-order approximation. We note that the current limit of the experimental measurement is about 1.0 meV. 

\begin{table}[htbp]
	\caption{\label{tab1} Exciton energies (meV) in monolayer WSe$_2$  encapsulated by hBN slabs, calculated by the regulated perturbation theory  and compared to the exact numerical solutions \cite{PhysRevB2023}.}
	\begin{ruledtabular}
		\begin{tabular}{l r r r r r }
				&	$E^{(0)}$	&	$E^{(2)}$	& $E^{(3)}$ &	$E_{\text{num}}$ \cite{PhysRevB2023} \\
			\hline
			$1s$	&	-164.98	&	-168.09	&	-168.55	&	-168.60		\\
$2s$	&	-39.46	&	-37.05	&	-38.52	&	-38.57	\\
$3s$	&	-16.94	&	-16.22	&	-16.40   	&	-16.56		\\
$4s$	&	-9.32	&	-9.03	&	-9.06   	&	-9.13		\\
$5s$	&	-5.87	&	-5.73	&	-5.74   	&	-5.77		\\
		\end{tabular}
	\end{ruledtabular}
\end{table}

We have also verified the accuracy of the RPT numerically for some other TMDC monolayers: WS$_2$, MoSe$_2$, and MoS$_2$. It shows the same high precision, less than $0.5$ meV, in the third-order approximation, which is not provided here but is available in Supplement \cite{Suppl}. All material parameters used in our calculations are provided in Table~\ref{tab3}.

\begin{table}[htbp]
	\caption{\label{tab3} Material paramaters: the dielectric constant $\kappa$, exciton reduced mass $\mu$, screening length $r_0$, 
effective Rydberg energy Ry$^*=\mu e^4/32\pi^2\varepsilon_0^2\kappa^2\hbar^2$, and effective Bohr radius $a_0^*=4\pi\varepsilon_0\kappa\hbar^2/\mu e^2$.}
	\begin{ruledtabular}
		\begin{tabular}{l r r r r r r }
				&	$\kappa$	&	$\mu$ ($m_e$)	& $r_0$ (nm) &	Ry$^*$ (meV) & $a^*_0$ (nm) & Refs.\\
			\hline
			WSe$_2$	&	4.34 &	0.19 &	4.21 &	137.24	&1.209	& \cite{PhysRevB2023}\\
WS$_2$	&		4.16&	0.175	&	3.76	&	137.59	&1.258 &\cite{PhysRevB2023}\\
MoSe$_2$ &	4.40	&	0.35	   &	 3.90  	&	245.97	&  0.665   &	\cite{NAT2019}\\
MoS$_2$	&	  4.45	&	0.275	&	 3.40 	&	188.94	&  0.856   &\cite{NAT2019}\\
MoTe$_2$ & 4.30 &0.37         &  6.10      &   272.26     &    0.615     & \cite{Suppl}\\

		\end{tabular}
	\end{ruledtabular}
\end{table}

\textit{Analytical exciton energies --} As shown above, the regulated perturbation theory for the corrections of up to the third-order approximation give extremely accurate exciton energies with precision within 0.5 meV compared to the numerically exact solutions. Even for the energies in the second-order approximation, the precision is about 1.0 meV, a limit for experimental measurement. Therefore, we consider them for analytical exciton energies. Indeed, by substituting the explicit forms of matrix elements $h_k$, $r_k$, $v_{jk}$, and $v^{\text{\tiny{\it R}}}_{jk}$ into RPT equations, we get particular exciton energies in the zero-order approximation as
\begin{eqnarray}\label{eq5}
E^{(0)}_{ns}=\frac{\kappa^2}{4{r_0^*}^2}\frac{1}{\lambda^2}-\frac{2\kappa}{r_0^*}f_{n}(\lambda).
\end{eqnarray}
Here, $r_0^*=r_0/a_0^*$ is defined as a dimensionless screening length; $\lambda$ is a variational parameter defined from the free parameter $\omega$ by  $\lambda=\kappa/\omega r_0^*$; $f_n (\lambda)$ is a function dependent on $\lambda$ only, whose explicit form is given in Supplement~\cite{Suppl}. For convenience, the principle quantum number is defined from now on as $n=1,2,3,\dots$ consistent with the well-known Rydberg series of a hydrogen atom. 

Formula \eqref{eq5} depends on parameter $\lambda$, which we consider a variational parameter and determine by the equation
\begin{eqnarray}\label{eq6}
\frac{\partial E^{(0)}_{ns}}{\partial \lambda}=0.
\end{eqnarray}
We must solve this equation to get $\lambda$ and plug it into Eq.~\eqref{eq5} to get the needed analytical energies. We demonstrate the solving process for $ns$ states ($n \leq 5$) in Supplement \cite{Suppl} and provide only the main idea and final results in the main text.

Because of the complexity of function $f_n(\lambda)$, analytically solving Eq.~\eqref{eq6} is not trivial, requiring the Taylor expansion of this function. From another side, numerically solving this equation for specific cases of monolayers WSe$_2$, WS$_2$, MoSe$_2$, and MoS$_2$ encapsulated by hBN slabs with the material parameters given in Table~\ref{tab3}, we get particular solutions for $1s$ state: $\lambda=0.803, 0.863, 0.582,$ and $0.750$, corresponding to each material. These results suggest we expand function $f_1(\lambda)$ around the average value $\lambda_0=0.75$. This expansion leads Eq.~\eqref{eq6} to a polynomial equation concerning an unknown $\delta \lambda = \lambda-0.75$,  making it easy to get approximate analytical solutions consistent with the numerical ones. Analogically, we get $\lambda$ for other $ns$ states by expanding functions $f_n (\lambda)$ around the points $\lambda_0=(n+0.5)/2$ and then solving Eq.~\eqref{eq6}. We combine all the solutions in one generalized formula as $\lambda=\lambda_0 +\delta\lambda$ with
\begin{eqnarray}\label{eq7}
\delta \lambda=\lambda-(n+0.5)/2=\left(0.47\, {\kappa}/{r_0^*}-0.525\right) n\,.
\end{eqnarray}

We note that the actual solutions of Eq.~\eqref{eq6} are not precisely as given by formula~\eqref{eq7}; however, a slight modification has been made to get a general formula for all states $ns$. This approach is reasonable because exact energies should not depend on the free parameter, and its modification around the extremum point does not change the solutions much. Reference \cite{Hoangbook2015} provides a detailed discussion of how to choose such a free parameter. 

Plugging $\lambda$ from Eq.~\eqref{eq7} into Eq.~\eqref{eq5}, we get exciton energies in analytical form, however, with more than 5\%  precision compared to the exact numerical solutions. So, for a higher precision, we will add the second-order correction of the perturbation theory to the energy \eqref{eq5}, which does not change its form. After some analytical transformations and reasonable truncations of the polynomials, whose details are provided in Supplement~\cite{Suppl}, we finally obtain exciton energy for the $ns$ state as
\begin{eqnarray}\label{eq8}
E_{ns}&=&-\frac{{\text{Ry}}^*\times P_n}{(n-0.5+0.479 \,r_0^*/\kappa)^2}.
\end{eqnarray}
Here, function $P_n$ has the form
\begin{eqnarray}\label{eq9}
P_n(r_0^*/\kappa)= \frac{\exp{(a_n\, \Delta)}}{1+b_n \,\Delta} 
\end{eqnarray}
with $\Delta=(0.47\, {\kappa}/{r_0^*}-0.525) $, coefficients 
$a_n=4.3/n-{6.6}/{n^2}+{3.2}/{n^3}$ and $b_n=5.3/n -{7.6}/{n^2}+{4.0}/{n^3}.$
$P_n$ is a function of variable $r_0^*/\kappa$ only with values around 1.0 and $P_n \rightarrow 1$ when $n \rightarrow +\infty$, so we can interpret it effective screening charge correction.

Table \ref{tab2} gives exciton energies for the $ns$ states ($n\leq 5$) calculated by the analytical formula \eqref{eq8} compared with the exact numerical solutions for monolayer TMDCs \cite{PhysRevB2023}. We note that Ref.~\cite{PhysRevB2023} provides only exciton energies for WSe$_2$ and WS$_2$; however, following the method given in this reference, we also calculate exact numerical exciton energies for MoSe$_2$ and MoS$_2$ to benchmark our analytical energies. The accuracy of the formula \eqref{eq8} is high, about 1.0 meV (an experimental precision limit) or less. 

\begin{table}[htbp]
	\caption{\label{tab2} Exciton energies of $ns$ states calculated by the analytical formula \eqref{eq8} compared with the numerical exact solutions in Ref.~\cite{PhysRevB2023} for monolayer TMDCs encapsulated by hBN.}
	\begin{ruledtabular}
		\begin{tabular}{l r r r r }
				&	WSe$_2$	&	WS$_2$  & MoSe$_2$ &	MoS$_2$ \\
			\hline
$r_0^*/\kappa $ &	0.803 &	0.719 &	1.333 &	0.893 \\
\hline
$E_{\text{1s}}$ (meV)	&	-167.96 &	-177.80 &	-229.88	&	-219.34 \\
Ref.~\cite{PhysRevB2023}  &	-168.60	&	 -178.62 &	 -231.96 &	-220.18 \\
\hline
$E_{\text{2s}}$ (meV)	&	-37.94 &	-39.12 &	-58.71	&	-50.83 \\
Ref.~\cite{PhysRevB2023}  &	-38.57	&	 -39.73 &	 -60.63 &	-51.81 \\
\hline
$E_{\text{3s}}$ (meV)	&	-16.28 &	-16.61 &	-26.60	&	-22.04 \\
Ref.~\cite{PhysRevB2023}  &	-16.56	&	 -16.90 &	 -27.31 &	-22.46 \\
\hline
$E_{\text{4s}}$ (meV)	&	-9.00 &	-9.13 &	-15.07	&	-12.25 \\
Ref.~\cite{PhysRevB2023}  &	-9.13	&	 -9.28 &	 -15.41 &	-12.44 \\
\hline
$E_{\text{5s}}$ (meV)	&	-5.70 &	-5.77 &	-9.68	&	-7.78 \\
Ref.~\cite{PhysRevB2023}  &	-5.77	&	 -5.85 &	 -9.87 &	-7.89 \\

		\end{tabular}
	\end{ruledtabular}
\end{table}

The formula \eqref{eq8} has a proper asymptotic behavior. When $n\rightarrow +\infty$, $P_n \rightarrow 1$ and  $n \gg 0.479\, r_0^*/\kappa$, so exciton energy decribed by Eq.~\eqref{eq8} becomes the 2D hydrogen atom energy $E_{ns} \rightarrow -{\text{Ry}^*}/(n-0.5)^2$.  Also, formula \eqref{eq8} complements the result of Molas {\it {et al.}} \cite{Molas2019} well. Indeed, Molas suggests the exciton energies of monolayer TMDCs by the ladder $E_{ns}=-$Ry$^*/(n+\delta)^2$, where Ry$^*$ and $\delta$ are from best fitting with experimental data. For monolayer WSe$_2$ encapsulated by hBN slabs, Ry$^*=140.5$ meV and $\delta=-0.083$. In our case, the effective Rydberg energy Ry$^*=137.2$ meV after a modification by the factor $P_n\sim 0.96\%$ (for $1s$ state) has the value 131.5 meV. Our theory gives a formula $E_n=-137.2\times P_n$ meV$/(n-0.115)^2 $ for exciton energies, which matches Molas's formula. This consistency also occurs for other monolayers such as WS$_2$, MoSe$_2$, and MoS$_2$. The difference in our theory is the effect of charge screening via the factor $P_n$, which is not considered in Molas's study.

\textit{Validation with the available experimental data --} To obtain analytical energies \eqref{eq8}, we expand the equation around the specific points of $\lambda$ with the condition $\lambda \gg \delta\lambda$, which defines the boundary of our theory, the working range of $r_0^*/\kappa$ as
\begin{equation}\label{eq10}
  0.46 \ll  r_0^*/\kappa \ll 18.8. 
\end{equation}
By definition, $r^*_0/\kappa=\mu/\kappa^2 \times r_0/a_0$, $a_0=0.0529$ nm - Bohr radius, so boundary \eqref{eq10} leads to a working range for $\kappa$, in particular for WSe$_2$, $0.89 \ll \kappa \ll 5.73$.

Furthermore, we verify formula \eqref{eq8} by comparing it with experimental data for monolayers WSe$_2$ \cite{PRL2018}, WS$_2$, MoSe$_2$, MoS$_2$, and MoTe$_2$ \cite{NAT2019}. The comparison in Fig.~\ref{fig1} demonstrates a high consistency between the experiments and theory.
For MoTe$_2$, material parameters are given in \cite{NAT2019} with a large uncertainty. However, we choose their values (in Table \ref{tab3}) so that our theory best fits the experimental data.

Experimental data of Liu {\it{et al.}} \cite{Liu2019} for monolayer WSe$_2$ also support our theory: $\Delta E_{2s-1s}=131$(130.3) meV, $\Delta E_{3s-2s}=21$ (21.7) meV, $\Delta E_{4s-2s}=30$ (29.0) meV, where the values in parenthesis are theoretical from Eq.~\eqref{eq8}.

\begin{figure}[t]
\begin{center}
\includegraphics[width=0.86 \columnwidth]{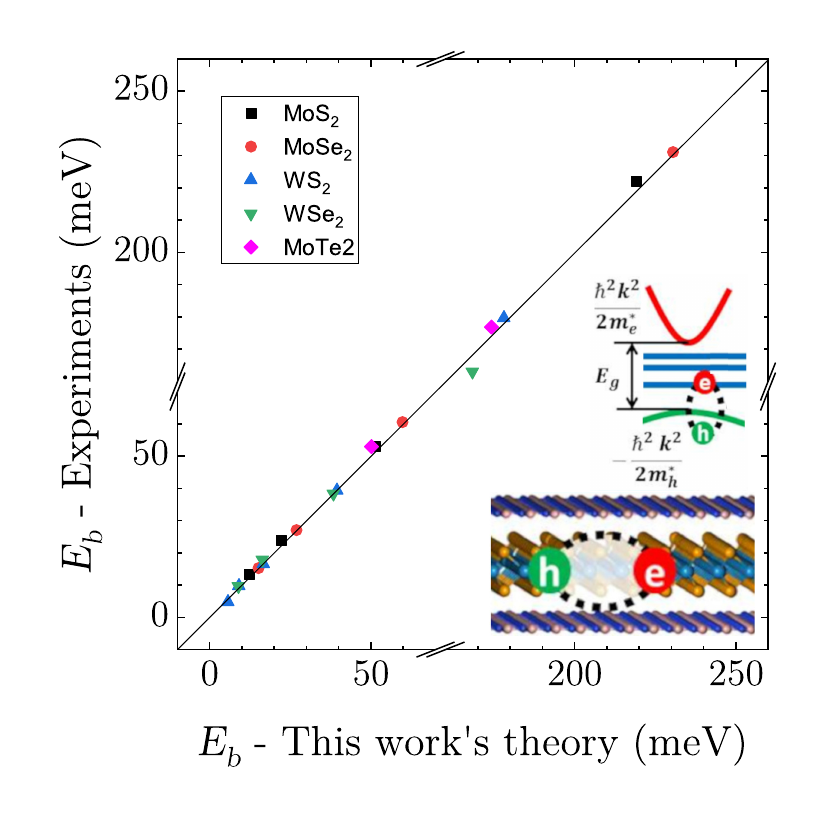}
\caption{Exciton binding energies $E_b$ for the $ns$ states in monolayers WSe$_2$, WS$_2$, MoSe$_2$, MoS$_2$, and MoTe$_2$ calculated by formula \eqref{eq8} (horizontal axis) compared with the experimental data \cite{NAT2019, PRL2018} (vertical axis). Material parameters used for calculations are provided in Table  \ref{tab3}. }
\label{fig1}
\end{center}
\end{figure} 

\textit{Conclusion --} A novel approach based on the perturbation theory has been developed, incorporating some elements of the FK operator method to achieve the energies of $s$-state excitons in monolayer transition-metal dichalcogenides. The third-order approximation of perturbation theory presents almost exact solutions, while the second-order approximation gives exciton energies with high precision, within 1.0 meV compared to the exact numerical solutions. This accuracy suggests we use them for analytical exciton energies. Although simple and universal for $s$-states, the constructed analytical formula describes experimental exciton energies for a wide range of monolayer TMDCs. This general approach could be applied to other 2D semiconduction materials.

\textit{Acknowledgments --} This work was funded by Vietnam Ministry of Education and Training under grant number B2022-SPS-09-VL and carried out by the high-performance cluster at Ho Chi Minh City University of Education, Vietnam.

\textit{Author contribution --} V.-H. Le conceptualized the work; V.-H. Le, H.T. Dinh, Dai-Nam Le, and N.-H. Phan performed analytical formulation, N.-H. Phan, Duy-Nhat Ly, N.-Q. Nguyen, and P.-T. Doan carried out numerical calculations; V.-H. Le, Dai-Nam Le, H.T. Dinh, and N.-T. D. Hoang analyzed data; V.-H. Le, Dai-Nam Le, and N.-H. Phan wrote and edited the manuscript.


\bibliography{refs}

\pagebreak
\widetext

\setcounter{section}{0}
\setcounter{equation}{0}
\setcounter{figure}{0}
\setcounter{table}{0}
\setcounter{page}{1}

\renewcommand{\thepage}{S-\arabic{page}} 
\renewcommand{\thesection}{S-\Roman{section}}  
\renewcommand{\thetable}{S-\Roman{table}}  
\renewcommand{\thefigure}{S-\arabic{figure}}
\renewcommand{\theequation}{S-\arabic{equation}}

\begin{center}
	\textbf{\large
		Supplemental materials for:\\Thermal effect on magnetoexciton energy spectra in monolayer \\transition-metal dichalcogenides
	}\\
	
	Hanh T. Dinh, $^{1,\dagger}$ Ngoc-Hung Phan,$^{1,\dagger}$ Duy-Nhat Ly,$^1$, Dai-Nam Le,$^2$, Ngoc-Tram D. Hoang,$^1$ Nhat-Quang Nguyen,$^1$ Phuoc-Thien Doan,$^1$ and Van-Hoang Le $^{1}$\\
	\textit{$^1$Computational Physics Key Laboratory K002, Department of Physics, Ho Chi Minh City University of Education, Ho Chi Minh City 72759, Vietnam\\
		$^2$ Department of Physics, University of South Florida, Tampa, Florida 33620, USA\\
		$^{\dagger}$These authors contributed equally to this work.
	}\\
	
	Emails: hoanglv@hcmue.edu.vn (V.-H. Le).\\
	
\end{center}

\maketitle

\section{\label{sec:S1} Analytical matrix elements }
\label{equation}

\subsection{\label{S1.1} The Schr\"{o}dinger equation for an exciton in monolayer TMDCs via the Levi-Civita transformation}

\textit{The Schr\"{o}dinger equation} for the relative motion of electron and hole can be written in atomic units as
\begin{eqnarray}\label{s1}
\left\{ - \left( \frac{\partial^2}{\partial x^2}+\frac{\partial^2}{\partial y^2}\right)  
     + {\hat V}_{h-e}(r)  -E\right\} \psi(x, y) = 0,
\end{eqnarray}
where $r=\sqrt{x^2+y^2}$; energy $E$ and coordinates $x, y$ are given in the effective Rydberg energy  Ry$^{*}=\mu e^4/16\pi^2\varepsilon_0^2\kappa^2\hbar^2$ and effective Bohr radius $a_0^{*}=4\pi\varepsilon_0\kappa\hbar^2/\mu e^2$ scaled by the exciton reduced mass $\mu=m^{*}_{e} m^{*}_{h}/(m^{*}_{e}+ m^{*}_{h})$ and dielectric constant $\kappa$ of the surounding medium; $m^{*}_{e}$ and $m^{*}_{h}$ are respectively the effective masses of the electron and hole.  Keldysh potetial has the form in atomic units as
\begin{equation}\label{s2}
{\hat V}_{e-h}(x,y)=-2 
\int\limits_{0}^{+\infty} \frac{dq}{ \sqrt{1+\alpha^2 q^2} }\;
 \textrm{e}^{-qr},
\end{equation}
where $\alpha=r_0/\kappa a_0^*$ with the screening length $r_0$. We note that most studies use the Keldysh potential expressed via Struve and Bessel functions, which are suitable only for numerical calculations.

\textit{The Levi-Civita transformation --} For more effectiveness in solving proccess, we apply the Levi-Civita transformation 
\begin{equation} \label{s3}
x=u^2-v^2,\; y=2uv,
\end{equation}
to the Schr{\"o}dinger equation \eqref{s1} to rewrite it in the $(u, v)$ space as
\begin{eqnarray}\label{s4}
\left\{ 
- \frac{1}{4} \left( \frac{\partial^2}{\partial {u}^2} +\frac{\partial^2}{\partial {v}^2}\right) 
+ \hat{V}_K(u, v)  - E  \left( {u}^2+{v}^2\right) \right\} \psi(u, v) = 0,
\end{eqnarray}
where the interaction potential in the $(u, v)$ space is defined as ${\hat V}_K(u, v)=\left( {u}^2+{v}^2\right) {\hat V}_{h-e}$, which has the form
\begin{equation}\label{s5}
{\hat V}_K(u,v)=- 2
\int\limits_{0}^{+\infty} \frac{dq}{ \sqrt{1+\alpha^2 q^2} }\;
 \textrm{e}^{-q(u^2+v^2)}(u^2+v^2).
\end{equation}
To obtain equations \eqref{s4} and for further use, we also derive the distance $r$ and the angular momentum $\hat l_z$ in $(u, v)$ coordinates as follows
\begin{eqnarray}\label{s6}
r=u^2+v^2,\quad
{\hat l_z}= -\frac{i}{2}\left( v\frac{\partial}{\partial u} - u\frac{\partial}{\partial v} \right).
\end{eqnarray}
For the application of the Levi-Civita transformation to two-dimensional atomic systems and calculation technique, one can also find more details in Ref.~\cite{giang1993}.

\subsection{\label{S1.2} Algebraic formalism with annihilation and creation operators} 
We rewrite the Schr\"{o}dinger equation \eqref{s4} in the form
\begin{equation}\label{s7}
\left( {\hat H} - E {\hat R} \right) |\psi\rangle=0,
\end{equation}
where
\begin{eqnarray}\label{s8}
{\hat H}= -\frac{1}{4} {\hat T} + {\hat V}_K,\quad
{\hat T}=\dfrac{\partial^2}{\partial {u}^2} +\dfrac{\partial^2}{\partial {v}^2}, \quad
{\hat R}={u}^2+{v}^2.
\end{eqnarray}
Equation \eqref{s7} is similar to those describing a two-dimensional anharmonic oscillator, suggesting we use its algebraic representation via annihilation and creation operators for convenience in calculations. Using the algebraic method, we then use the harmonic oscillator's wave functions to establish a basis set and calculate all necessary matrix elements. More details of this algebraic approach can be found in monograph~\cite{Hoangbook2015}. 

First, we define annihilation and creation operators as follows
\begin{eqnarray}\label{s9}
\hat \alpha  = \sqrt {\frac{\omega }{2}} \left( {u + \frac{1}{\omega }\frac{\partial }{{\partial u}}} \right),
\quad \hat \alpha ^ +  = \sqrt {\frac{\omega }{2}} \left( {u - \frac{1}{\omega }\frac{\partial }{{\partial u}}} \right), \nonumber \\
\quad \hat \beta  = \sqrt {\frac{\omega }{2}} \left( {v + \frac{1}{\omega }\frac{\partial }{{\partial v}}} \right),
\quad \hat \beta ^ +  = \sqrt {\frac{\omega }{2}} \left( {v - \frac{1}{\omega }\frac{\partial }{{\partial v}}} \right).
\end{eqnarray}
Here, $\omega$ in the above definition is usually an angular frequency of the harmonic oscillator. However, we can freely choose its value to manipulate the converge rate of the solutions, see \cite{Hoangbook2015} and references therein.
Then, we use a canonical transformation to obtain operators
\begin{eqnarray}\label{s10}
\hat a = \frac{1}{{\sqrt 2 }}\left( {\hat \alpha  - i\hat \beta } \right),
\quad \hat a^ + = \frac{1}{{\sqrt 2 }}\left( {{{\hat \alpha }^ + } + i{{\hat \beta }^ + }} \right),
\quad \hat b = \frac{1}{{\sqrt 2 }}\left( {\hat \alpha  + i\hat \beta } \right),
\quad \hat b^ + = \frac{1}{{\sqrt 2 }}\left( {{{\hat \alpha }^ + } - i{{\hat \beta }^ + }} \right).
\end{eqnarray}
Combinning two equations \eqref{s9} and \eqref{s10}, the definition for the annihilation and creation operators is as follows
\begin{eqnarray}\label{s11}
{\hat a} &=& \frac{\sqrt{\omega}}{2}  \left( u -i\,v \right) 
        + \frac{1}{2\,\sqrt{\omega}} \left( \frac{\partial}{\partial u} - i\, \frac{\partial}{\partial v} \right), \nonumber \\
{\hat a}^{+} &=& \frac{\sqrt{\omega}}{2}  \left( u +i\,v \right) 
        - \frac{1}{2\,\sqrt{\omega}} \left( \frac{\partial}{\partial u} + i\, \frac{\partial}{\partial v} \right), \nonumber \\
{\hat b} &=& \frac{\sqrt{\omega}}{2}  \left( u +i\,v \right) 
        + \frac{1}{2\,\sqrt{\omega}} \left( \frac{\partial}{\partial u} + i\, \frac{\partial}{\partial v} \right), \nonumber \\
{\hat b}^{+} &=& \frac{\sqrt{\omega}}{2}  \left( u -i\,v \right) 
        - \frac{1}{2\,\sqrt{\omega}} \left( \frac{\partial}{\partial u} - i\, \frac{\partial}{\partial v} \right).
\end{eqnarray}

The purpose of definition \eqref{s11} for operators $\hat a$, $\hat a^{+}$, $\hat b$, and $\hat b^{+}$ is twofold. First, these operators remain satisfying the commutation relations 
\begin{equation}\label{s12}
	\left[{\hat a}, {\hat a}^{+} \right]=1,\;\;\left[{\hat b}, {\hat b}^{+} \right]=1,
\end{equation} 
typical for annihilation and creation operators; and second, the angular momentum becomes a neutral operator as in the form
\begin{equation}\label{s13}
{\hat l_z}=\frac{1}{2}( {\hat a}^{+}{\hat a}- {\hat b}^{+}{\hat b}).
\end{equation}
We also note that, operators ${\hat a}$, ${\hat a}^{+}$ are independent from operators ${\hat b}$, ${\hat b}^{+}$, meaning they commute with each other as
\begin{equation}\label{s14}
	\left[{\hat a}, {\hat b} \right]=\left[{\hat a}^{+}, {\hat b}^{+} \right]=
     \left[{\hat a}, {\hat b}^{+} \right]=\left[{\hat a}^{+}, {\hat b} \right]=0.
\end{equation}

Finally, using the annihilation and creation operators defined by equations \eqref{s11}, we can rewrite all the terms in the Schr{\"o}dinger equation (\ref{s7}) as
\begin{eqnarray}\label{s15}
{\hat T}=&\dfrac{\partial^2}{\partial {u}^2} +\dfrac{\partial^2}{\partial {v}^2} &= \;\omega\left({\hat a}{\hat b} 
    + {{\hat a}}^{+}{{\hat b}}^{+} - {{\hat a}}^{+}{\hat a} - {{\hat b}}^{+}{\hat b}  - 1 \right),\nonumber\\
{\hat R}=&{u}^2+{v}^2 &= \;\frac{1}{\omega} \left( {\hat a}{\hat b} + {{\hat a}}^{+} {{\hat b}}^{+} + {{\hat a}}^{+}{\hat a} + {{\hat b}}^{+}{\hat b} + 1 \right).
\end{eqnarray}
The potential operator $\hat V_K$ can be expressed via $\hat R$ as
\begin{equation}\label{s16}
{\hat V}_K=- 2
\int\limits_{0}^{+\infty} \frac{dq}{ \sqrt{1+\alpha^2 q^2} }\;
 \textrm{e}^{-q \hat R}{\hat R}.
\end{equation}

\subsection{\label{S1.3} Harmonic oscillator basic set and algebraically calculating matrix elements }
\textit{Basis set --}  The Hamiltonian of two-dimentional isotropic harmonic oscillator associated with the annihilation and creation operators \eqref{s11} has the algebraic form $\hat H_{\text{osc}}=\omega ({\hat a}^{+}{\hat a}+{\hat b}^{+}{\hat b}+1)$. We will choose the eigenvectors of this Hamiltonian to be a basis set for our resolving process considering the Schr{\"o}dinger equation (\ref{s4}) describing a two-dimensional anharmonic oscillator in the $(u, v)$- space. In addition, because the angular momentum conservation of excitons in monolayer TMDCs, we also require the basis set functions to be eigenfunctions of the operator ${\hat l}_z$ (Eq.\eqref{s13}). As a results, we obtain 
\begin{equation}\label{s17}
          {| k,m \rangle} = \frac{1}{\sqrt{{(k+m)}! {(k-m)}!}} 
            ({\hat a}^{+})^{k+m} ({\hat b}^{+})^{k-m}| 0 (\omega) \rangle
\end{equation}
as a basis set, where the vacuum state $|0 (\omega)\rangle$ satisfies the equations:
\begin{equation}\label{s18}
       {\hat a}\,{| 0 (\omega) \rangle} = 0,\;\; {\hat b} \,| 0 (\omega) \rangle =0.
\end{equation}
Considering the algebraic form \eqref{s13}, we can verify that ${\hat l}_z\, {| k,m \rangle}=m\, {| k,m \rangle}$ with $m=0, \pm 1, \pm 2, ...$. Therefore the running index $m$ of the basis set is fixed by the magnetic quantum number of the system. The other running index $k$ is an integer with the values $k \geq |m|$. By using the commutation relations \eqref{s12}, \eqref{s14} and equations \eqref{s18}, we can also certify the orthonormality of the basis set functions (\ref{s17}), i.e. 
$\langle k, m| k', m'\rangle =\delta_{k, k'}\delta_{m, m'},$
where $\delta_{ij}$ is the Kronecker delta. 

\textit{Matrix elements -- } 
Commutation relations \eqref{s12}, \eqref{s14}, and equations \eqref{s18} are useful for constructing algebraic calculation techniques based on the annihilation and creation operators.  This kind of algebraic technique is given in textbooks on quantum mechanics and monograph \cite{Hoangbook2015}. For all the operators in the Schr{\"o}dinger equation (\ref{s4}), matrix elements with respect to the basis set \eqref{s17} can be calculated and formulated analytically in Ref.~\cite{PhysRevB2023}.  In the present paper, we consider only $s$-states ($m=0$) so the basis set becomes
\begin{equation}\label{s19}
          {| n \rangle} = \frac{1}{n!} 
            ({\hat a}^{+}{\hat b}^{+})^{n}| 0 (\omega) \rangle.
\end{equation}

(i) It is easy to derive the following equations
\begin{eqnarray}\label{s20}
   ( {\hat a}^{+} {\hat a} + {\hat b}^{+} {\hat b} +1)\, {|n \rangle} 
         = (2n+1)\, {|n \rangle}, \quad
   {\hat a}^{+} {\hat b}^{+} {|n \rangle} 
       = {(n+1)}\, {|n+1 \rangle}, \quad
   {\hat a}{\hat b}\; {|n \rangle} = {n} \;{|n-1\rangle},
\end{eqnarray}
which lead to the matrix elements
\begin{eqnarray}\label{s21}
&& N_{jk} = \langle j|\, ({\hat a}^{+} {\hat a} + {\hat b}^{+} {\hat b}+1) \, {|k\rangle}= (2k+1)\, \delta_{jk}, \nonumber\\
&& M_{jk}^{+} = \langle j| \,{\hat a}^{+} {\hat b}^{+}\, {|k\rangle} = {(k+1)} \,\delta_{j, k+1}, \nonumber\\
&& M_{jk} = \langle j|\, {\hat a}{\hat b}\, {|k\rangle} = {k} \,\delta_{j, k-1},
\end{eqnarray}
useful for calculating matrix elements of the operators $\hat T$ and $\hat R$. We obtain
\begin{eqnarray}\label{s22}
\omega {R}_{jk}&=&\omega \,\langle j|\, {\hat R} \, {|k\rangle}= N_{jk} + M_{jk}+ M^{+}_{jk}\qquad\nonumber\\
&=& (2k+1)\, \delta_{jk}+ {k} \,\delta_{j, k-1}+ {(k+1)} \,\delta_{j, k+1}\,,
\end{eqnarray}
\begin{eqnarray}\label{s23}
\frac{1}{\omega}{T}_{jk}&=& \frac{1}{\omega}\,\langle j|\, {\hat T} \, {|k\rangle}= -N_{jk} + M_{jk}+ M^{+}_{jk}\nonumber\\
&=& -(2k+1)\, \delta_{jk}+ {k} \,\delta_{j, k-1}+ {(k+1)} \,\delta_{j, k+1}\,.
\end{eqnarray}

(ii) Differently, it is not trivial to calculate matrix elements of the operator $\hat V_K$. However, we can do it by using the technique of constructing operators in a normal form of annihilation and creation operators, given in Ref.~\cite{Hoangbook2015}. Indeed, using the formulae (\ref{s16}), we have
\begin{eqnarray}\label{s24}
      V^{\text{\tiny K}}_{jk} &=& \langle j|{\hat V_K}{| k \rangle}
    = (2k+1)\, U_{jk} +{k}\, U_{j, k-1} +{(k+1)}\, U_{j, k+1},
\end{eqnarray}
where
\begin{eqnarray}\label{s25}
    U_{jk} &=& -2 \int\limits_{0}^{+\infty} \frac{dq}{\sqrt{1+{\omega}^2{\alpha}^2 q^2}}
\times \; \langle j|\, \textrm{e}^{-q\, (\hat{N}+\hat{M}+\hat{M}^{+})} |k\rangle.
\end{eqnarray}
Here, we use the notations $\hat N= {\hat a}^{+}{\hat a}+{\hat b}^{+}{\hat b}+1$, $\hat M= {\hat a}{\hat b}$, and ${\hat M}^{+}={\hat a}^{+}{\hat b}^{+}$.
To calculate matrix elements $U_{jk}$, we need the operator $\hat{O}=\textrm{e}^{-q\,(\hat{N}+\hat{M}+\hat{M}^{+})}$ in the normal form, in which the annihilation operator $\hat M$ in the right, the neutral operator $\hat N$ in the middle, and the creation operator ${\hat M}^{+}$ in the left. For operators creating a close algebra, there exists a procedure to create the normal form of the exponential operator \cite{Hoangbook2015}. In our case, we have the communication relations
\begin{equation}\label{s26}
 \left[\hat{M}, \hat{N} \right]= 2 \hat{M},\quad \left[\hat{N}, \hat{M}^{+} \right]= 2\hat{M}^{+},\quad
         \left[\hat{M}, \hat{M}^{+} \right]= \hat{N}, 
\end{equation}
that means operators $\hat N$, $\hat M$, and $\hat M^+$ creat a close algebra. Following this procedure given on page 232--233 of Ref. \cite{Hoangbook2015}, we have the formula
\begin{eqnarray}\label{s27}
\hat{O}&=&\textrm{e}^{-q \,(\hat{N}+\hat{M}+\hat{M}^{+})}
= e^{-\frac{q}{1+q}\,\hat{M}^{+}} 
               e^{-\ln(1+q)\, \hat{N}} e^{-\frac{q}{1+q}\,\hat{M}}.
\end{eqnarray}
Then, expanding the exponential operators in Eq.\eqref{s27} by equation $e^X=\sum_{s=0}^{+\infty}\frac{1}{s!}X^s$ and using the fomulae obtained from Eqs.(\ref{s20}) as
\begin{eqnarray}\label{s28}
  ({\hat M}^{+})^s {|k \rangle} =  \frac{(k+s)!}{k!} \, {|k+s \rangle}, \quad
   {\hat M}^s\, {|k \rangle} = \frac{k!}{(k-s)!} \;{|k-s \rangle},\quad
 {\hat N}^s\, {|k \rangle} = (2k+1)^s\, {|k \rangle}, 
\end{eqnarray}
we have
\begin{eqnarray}\label{s29}
 O_{jk}= \langle j | \textrm{e}^{-q\,(\hat{N}+\hat{M}+\hat{M}^{+})} {|k\rangle} 
 =(-1)^{j+k} \sum_{s=0}^{\text{min}(k,j)}  
          {{j}\choose{s}}
   {{k}\choose {s}}   \frac{q^{j+k-2s}}{(1+q)^{j+k+1}},
 \end{eqnarray}
where ${n \choose k}$ is the binomial coefficient, i.e. ${ n \choose k}= \frac{n!}{(n-k)! k!}$.

Plugging Eq.(\ref{s29}) into (\ref{s25}), we obtain the final result for $U_{jk}$ as
\begin{eqnarray}\label{s30}
 U_{jk} &=& -\frac{2}{\omega\,\alpha} \sum_{s=0}^{\text{min}(k,j)} \sum_{t=0}^{j+k-2s} (-1)^{j+k+t}
    {{j+k-2s} \choose {t}} {{j}\choose{s}} {{k}\choose {s}} \nonumber\\
&&\quad\qquad\qquad\qquad\qquad\quad\times \int\limits_{0}^{+\infty} \frac{dq}{(1+q)^{2s+t+1}\sqrt{q^2+1/\omega^2\alpha^2}}.
\end{eqnarray}
In Eq. (\ref{s30}) the definite integral 
\begin{equation}\label{s31}
J_p (\lambda)=\int\limits_{0}^{+\infty} \frac{dq}{(1+q)^p\sqrt{q^2+\lambda^2}}
\end{equation}
with $p \geq 1$ and $\lambda=1/\omega\alpha>0$ is easy to calculate numerically. Besides, we can derive an iterative formula for this integral as follows:
\begin{equation}\label{s32}
J_p=\frac{(2p-3)J_{p-1}-(p-2)J_{p-2}+\lambda}{(\lambda^2+1)(p-1)}
\end{equation}
with $p \geq 2$, where $J_1(\lambda)$ has the following explicit formula:
\begin{equation}\label{s33}
J_1 (\lambda)=\frac{\ln{\left(\lambda+\sqrt{\lambda^2+1}\right)}+\ln{\left(1+\sqrt{\lambda^2+1}\right)}-\ln(\lambda) }{\sqrt{\lambda^2+1}}.
\end{equation}
Noting that althought $J_0(\lambda)$ is disvergent, relation \eqref{s32} is still valid for $p=2$ by considering the limit
\[ {\textrm{lim}}_{p \rightarrow 0} \;pJ_p(\lambda)=1.\]
so that
\begin{equation}\label{s34}
J_2(\lambda)=\frac{J_1(\lambda)-1+\lambda}{\lambda^2+1}.
\end{equation}

For solving the Schr\"{o}dinger equation \eqref{s7} with the basis set \eqref{s19}, we need the following matrix elements:
\begin{equation}\label{s35}
h_j=\langle j| \hat H|j\rangle,\quad v_{jk}=\langle j| \hat H| k\rangle\; (j\neq k),\quad 
r_j=\langle j| \hat R|j\rangle,\quad v^{\text{\tiny R}}_{jk}=\langle j| \hat R| k\rangle\; (j\neq k).
\end{equation}
From the above formulation, we can obtain these matrix elements explicitly as follows.
\begin{eqnarray}\label{s36}
h_j&=&\langle j| \hat H|j\rangle=-\frac{1}{4}T_{jj}+ V^{\text{\tiny K}}_{jj}
= \frac{\omega}{4} (2j+1)-\frac{2\kappa}{ \omega r_0^*}\left[(2j+1) u_{jj}+j u_{j,j-1}+ (j+1) u_{j,j+1}\right],\nonumber\\
v_{jk}&=&\langle j| \hat H|k\rangle  =-\frac{1}{4}T_{jk}+ V^{\text{\tiny K}}_{jk}\quad (j\neq k)\nonumber\\
&=& \frac{\omega}{4}\left( k \delta_{j,k-1}+j\delta_{j,k+1}\right)-\frac{2\kappa}{\omega r_0^*}\left[ (2k+1) u_{j,k}+k u_{j,k-1}+(k+1)u_{j,k+1}\right],
\end{eqnarray}
\begin{eqnarray}\label{s37}
r_j&=&\langle j| \hat R|j\rangle=R_{jj}
= \frac{1}{\omega}(2j+1),\nonumber\\
v^{\text{\tiny R}}_{jk}&=&\langle j| \hat R|k\rangle  =R_{jk}\quad (j\neq k)
= \frac{1}{\omega}\left( k \delta_{j,k-1}+j\delta_{j,k+1}\right),
\end{eqnarray}
where $r_0^*=r_0/a_0^*$ is a dimensionless screening length;
\begin{eqnarray}\label{s38}
 u_{jj}(\lambda) &=&  \sum_{s=0}^{j} \sum_{t=0}^{2j-2s} (-1)^{t}
    {{2j-2s} \choose {t}} {{j}\choose{s}}^2  J_{2s+t+1}(\lambda),
\end{eqnarray}
\begin{eqnarray}\label{s39}
 u_{jk}(\lambda) &=&  \sum_{s=0}^{\text{min}(k,j)} \sum_{t=0}^{j+k-2s} (-1)^{j+k+t}
    {{j+k-2s} \choose {t}} {{j}\choose{s}} {{k}\choose {s}} J_{2s+t+1}(\lambda).
\end{eqnarray}

\section{\label{sec:S2} Regulated perturbation theory of solving the Schr\"{o}dinger equation}
\label{FKMethod}
\subsection{\label{sec:S2A} Basic formulation of the perturbation theory}
The Schr\"{o}dinger equation considered in this study, Eq.~\eqref{s7}, has the form
\begin{equation}\label{s40}
\left( {\hat H} - E {\hat R} \right) \psi(u,v)=0,
\end{equation}
where operators $\hat H$ and $\hat R$ are hermite. Noting that this equation was obtained from the Schr\"{o}dinger equation in ($x, y$) space by the Levi-Civita transformation. 

We solve equation \eqref{s40} by the perturbation theory, deviding the operators as 
\begin{eqnarray}\label{s41}
\hat H = \hat H_0 +\beta\, \hat V,\quad
\hat R = \hat R_0 +\beta \hat V_R,
\end{eqnarray}
with the perturbation parameter $\beta$. As shown in the following calculations, this parameter is introduced formally and does not effect the final results. The operators $\hat H_0$ and $\hat R_0$ are chosen so that they can be considered zero-order approximation of operators $\hat H$ and $\hat R$  and have common solutions $\psi^{(0)}$, $E^{(0)}$. Following the idea of the Feranchuk-Komarov (FK) operator method, we choose  $\hat H_0$ and $\hat R_0$ so that their algebraic forms contain only neutral operators $\hat a^+ \hat a$ and $\hat b^+ \hat b$. It means these operators are diagonal respect to the basis set $|k\rangle$ (\eqref{s18}). The remain parts, $\hat V$ and $\hat V_R$, are non-diagonal, containing $\hat a^+$, $\hat a$, $\hat b^+$, and $\hat b$ evenly. In this case, the matrix elements of theses operators with respect to the basis set \eqref{s19} have the following properties
\begin{eqnarray} \label{s42}
&&\langle j| \hat H_0 |j\rangle=\langle j| \hat H |j\rangle=h_j , \quad  \langle j| \hat R_0 |j\rangle=\langle j| \hat R |j\rangle=r_j  \nonumber\\
&&\langle j| \hat V |k\rangle=\langle j| \hat H |k\rangle=v_{jk},
 \quad  \langle j| \hat V_R |k\rangle=\langle j| \hat R |k\rangle=v^{\text{\tiny R}}_{jk},\quad (j \neq k),
\end{eqnarray}
whose explicit expressions are given in Eqs.\eqref{s36} and \eqref{s37}.

Wave function $\psi$ and energy $E$ are found in the expansions: 
\begin{eqnarray}\label{s43}
\psi=\psi^{(0)}+\beta\Delta \psi^{(1)}+\beta^2\Delta \psi^{(2)}+\cdots=\psi^{(0)}+\sum_{s=1}^{+\infty} \beta^s\Delta\psi^{(s)}\nonumber\\
E=E^{(0)}+\beta\Delta E^{(1)}+\beta^2\Delta E^{(2)}+\cdots=E^{(0)}+\sum_{s=1}^{+\infty}\beta^s\Delta E^{(s)}\;.
\end{eqnarray}
Here, $\Delta \psi^{(s)}$ and $\Delta E^{(s)}$ are corrections on the wave function and energy in the $s$-order approximation.  

Plugging expansions \eqref{s43} into the Schr\"{o}dinger equation \eqref{s40}, we obtain 
\begin{eqnarray}\label{s44}
&&\left(\hat H_0 - E^{(0)} \hat R_0\right) \psi^{(0)}+ \beta \left[\left(\hat H_0 - E^{(0)} \hat R_0\right) \Delta\psi^{(1)}
                        +\left(\hat V -E^{(0)} \hat V_R- \Delta E^{(1)} \hat R_0\right) \psi^{(0)}    \right]\nonumber\\
&&\quad+\beta^2 \left[\left(\hat H_0 - E^{(0)} \hat R_0\right) \Delta\psi^{(2)}
                        +\left(\hat V -E^{(0)} \hat V_R- \Delta E^{(1)} \hat R_0\right) \Delta\psi^{(1)} 
                            -\left( \Delta E^{(1)} \hat V_R +\Delta E^{(2)} \hat R_0\right)\psi^{(0)}   \right]\nonumber\\
&&\qquad+\sum_{s=3}^{+\infty}\beta^s \left[\left(\hat H_0 - E^{(0)} \hat R_0\right) \Delta\psi^{(s)}
          + \left(\hat V -E^{(0)} \hat V_R- \Delta E^{(1)} \hat R_0\right) \Delta\psi^{(s-1)}\right.\nonumber\\
&&\quad\quad\qquad\left. -\sum_{t=1}^{s-2} \left( \Delta E^{(s-t)} \hat R_0 + \Delta E^{(s-t-1)} \hat V_R\right) \Delta \psi^{(t)}
- \left(\Delta E^{(s)} \hat R_0 +\Delta E^{(s-1)} \hat V_R\right) \psi^{(0)} \right]=0.
\end{eqnarray}

From Eq.~\eqref{s44}, we can get the equation for the zero-order approximation as
\begin{eqnarray}\label{s45}
\left(\hat H_0 - E^{(0)} \hat R_0\right) \psi^{(0)}=0.
\end{eqnarray}  
Because of the diagonality of the operators $\hat H_0 (\hat a^+ \hat a, \hat b^+ \hat b)$ and $\hat R_0 (\hat a^+ \hat a, \hat b^+ \hat b)$, the eigen vectors and eigen values of Eq.~\eqref{s45} are as follows
 \begin{eqnarray}\label{s46}
 \psi^{(0)}_n=|n\rangle, \quad E^{(0)}_n=\frac{h_n}{r_n}.
\end{eqnarray}  

We can also get the equation for corrections in second-order approximation as
\begin{eqnarray}\label{s47}
\left(\hat H_0 - E_n^{(0)} \hat R_0\right) \Delta\psi^{(1)}_n
                        +\left(\hat V -E_n^{(0)} \hat V_R- \Delta E^{(1)}_n \hat R_0\right) |n\rangle=0.
\end{eqnarray}
 Multiflying the conjugate vector $|n\rangle $ and integrating over the space, we get
\begin{eqnarray}\label{s48}
\langle n | \left(\hat H_0 - E_n^{(0)} \hat R_0\right) |\Delta\psi^{(1)}_n\rangle
                        +\langle n |\left(\hat V -E^{(0)}_n \hat V_R- \Delta E^{(1)}_n \hat R_0\right) |n\rangle=0.
\end{eqnarray}
Because of the hermite property, we rewrite the first term in Eq.~\eqref{s48} as  $ \langle \Delta\psi^{(1)}_n | (\hat H_0 - E^{(0)}_n)| n \rangle$ which means this first term vanishes. In other hand, from properties of matrix elements Eqs.~\eqref{s42}, we have $\langle n| \hat V |n\rangle =0$ and  $\langle n| \hat V_R |n\rangle =0$. As a consequence, Eq.~\eqref{s48} leads to $r_n \Delta E^{(1)}_n=0$. i.e., 
\begin{equation} \label{s49}
\Delta E^{(1)}_n=0\,.
\end{equation}
Furthermore, Eq.~\eqref{s47} can be used to determine the correction to the wave function $\Delta \psi^{(1)}_n$. First, dealing with the second term in the equation, we have $\hat V |n \rangle=\sum_{j=0, j\neq n}^{+\infty} v_{jn} |j\rangle $ , $\hat V_R |n \rangle=\sum_{j=0, j\neq n}^{+\infty} v^{\text{\tiny R}}_{jn} |j\rangle $. Plugging these equation into Eq.~\eqref{s47}, we obtain the following
\begin{eqnarray}\label{s50}
 \Delta\psi^{(1)}_n
                        =-\sum_{\begin{smallmatrix} j=0\\ j\neq n \end{smallmatrix}}^{+\infty} \frac{ v_{jn}-E^{(0)}_n v^{\text{\tiny R}}_{jn}}
                                  {h_j- E^{(0)}_n r_j}\, |j\rangle
=-\sum_{\begin{smallmatrix} j=0\\ j\neq n \end{smallmatrix}}^{+\infty} \frac{ r_n v_{jn}- h_n v^{\text{\tiny R}}_{jn}}
                                  {r_n h_j- h_n r_j}\, |j\rangle\,.
\end{eqnarray}

By this wav, we can get corrections to energies and wave functions for the second-order approximation as follows
\begin{eqnarray}\label{s51}
 \Delta E^{(2)}_n
                        =-\sum_{\begin{smallmatrix} j=0\\ j\neq n \end{smallmatrix}}^{+\infty} \frac{ \left( v_{nj}- E^{(0)}_n v^{\text{\tiny R}}_{nj}\right)  \left( v_{jn}- E^{(0)}_n v^{\text{\tiny R}}_{jn}\right)     }
                                  {\left(h_j- E^{(0)}_n r_j\right) {r_n}}
=-\sum_{\begin{smallmatrix} j=0\\ j\neq n \end{smallmatrix}}^{+\infty} \frac{ \left( r_n v_{nj}- h_n v^{\text{\tiny R}}_{nj}\right)
      \left( r_n v_{jn}- h_n v^{\text{\tiny R}}_{jn}\right)     }
                                 {\left(r_n h_j- h_n r_j\right)  {r_n}^2}\,,
\end{eqnarray}
\begin{eqnarray}\label{s52}
 \Delta \psi^{(2)}_n
                        &=&\sum_{\begin{smallmatrix} k=0\\ k\neq n \end{smallmatrix}}^{+\infty}
        \sum_{\begin{smallmatrix} j=0\\ j\neq k\\j\neq n \end{smallmatrix}}^{+\infty} 
                 \frac{ \left( v_{jn}- E^{(0)}_n v^{\text{\tiny R}}_{jn}\right)  \left( v_{kj}- E^{(0)}_n v^{\text{\tiny R}}_{kj}\right)     }
                                  {\left(h_j- E^{(0)}_n r_j\right)\left(h_k- E^{(0)}_n r_k\right) }\,|k\rangle\nonumber\\
&=&\sum_{\begin{smallmatrix} k=0\\ k\neq n \end{smallmatrix}}^{+\infty}
        \sum_{\begin{smallmatrix} j=0\\ j\neq k\\j\neq n \end{smallmatrix}}^{+\infty} 
          \frac{ \left( r_n v_{jn}- h_n v^{\text{\tiny R}}_{jn}\right) \left( r_n v_{kj}- h_n v^{\text{\tiny R}}_{kj}\right)     }
                                 {\left(r_n h_j- h_n r_j\right)\left(r_n h_k- h_n r_k\right)  }\,| k\rangle.
\end{eqnarray}

Considering Eq.~\eqref{s44}, we can obtain corrections to energies and wave functions in any order of approximation. However, we are interested only corrections to energies in the approximation of up to the third order as
\begin{eqnarray}\label{s53}
 \Delta E^{(3)}_n
                        &=&\sum_{\begin{smallmatrix} k=0\\ k\neq n \end{smallmatrix}}^{+\infty}
        \sum_{\begin{smallmatrix} j=0\\ j\neq k\\j\neq n \end{smallmatrix}}^{+\infty}  
           \frac{ \left( v_{nk}- E^{(0)}_n v^{\text{\tiny R}}_{nk}\right)  \left( v_{kj}- E^{(0)}_n v^{\text{\tiny R}}_{kj}\right) 
                     \left( v_{jn}- E^{(0)}_n v^{\text{\tiny R}}_{jn}\right)    }
                                  {\left(h_j- E^{(0)}_n r_j\right)\left(h_k- E^{(0)}_n r_k\right) {r_n}}\nonumber\\
&=&\sum_{\begin{smallmatrix} k=0\\ k\neq n \end{smallmatrix}}^{+\infty}
        \sum_{\begin{smallmatrix} j=0\\ j\neq k\\j\neq n \end{smallmatrix}}^{+\infty}  
           \frac{ \left( r_n v_{nk}- h_n v^{\text{\tiny R}}_{nk}\right)  \left( r_n v_{kj}- h_n v^{\text{\tiny R}}_{kj}\right) 
                     \left( r_n v_{jn}- h_n v^{\text{\tiny R}}_{jn}\right)    }
                                  {\left(r_n h_j- h_n r_j\right)\left(r_n h_k- h_n r_k\right) {r_n}^2}\,.
\end{eqnarray}

\subsection{\label{sec:S2B} Numerical results}
The analytical formulae above by the perturbation theory, Eqs.~\eqref{s46}, can be used to obtain the $s$-state energies in the zero-order approximation. Moreover, the corrections to energies in the approximation of up to third order can be calculated by Eqs.~\eqref{s49}, \eqref{s51}, and \eqref{s53}. We note that all the matrix elements in these formulae are given analytically by Eqs.~\eqref{s36} and \eqref{s37}. We will calculate energies for $1s$, $2s$, $3s$, $4s$, and $5s$ for monolayer TMDCs (WSe$_2$, WS$_2$, MoSe$_2$, and MoS$_2$) whose material parameters are given in Table~\ref{tabS1}.

\begin{table}[htbp]
	\caption{\label{tabS1} Material paramaters: the dielectric constant $\kappa$, exciton reduced mass $\mu$, screening length $r_0$, 
effective Rydberg energy Ry$^*=\mu e^4/32\pi^2\varepsilon_0^2\kappa^2\hbar^2$, and effective Bohr radius $a_0^*=4\pi\varepsilon_0\kappa\hbar^2/\mu e^2$.}
	\begin{ruledtabular}
		\begin{tabular}{l r r r r r r }
				&	$\kappa$	&	$\mu$ ($m_e$)	& $r_0$ (nm) &	Ry$^*$ (meV) & $a^*_0$ (nm) & Refs.\\
			\hline
			WSe$_2$	&	4.34 &	0.19 &	4.21 &	137.24	&1.211	& \cite{PhysRevB2023}\\
WS$_2$	&		4.16&	0.175	&	3.76	&	137.59	&1.256 &\cite{PhysRevB2023}\\
MoSe$_2$ &	4.40	&	0.35	   &	 3.90  	&	245.97	&  0.664   &	\cite{NAT2019}\\
MoS$_2$	&	  4.45	&	0.275	&	 3.40 	&	188.94	&  0.854   &\cite{NAT2019}\\

		\end{tabular}
	\end{ruledtabular}
\end{table}

We have  the s-state energies in zero-order approximation by Eq.~\eqref{s45} as
\begin{eqnarray}\label{s54}
  E^{(0)}_{\text{ns}}=\frac{h_{n-1}}{r_{n-1}},
\end{eqnarray}  
where $n=1,2,3,4,5$ cosresponding to the states  $1s, 2s, 3s, 4s, 5s$. Using the analytical formulae \eqref{s36} and \eqref{s37} for $h_j$ and $r_j$, we get
\begin{eqnarray}\label{s55}
  E^{(0)}_{\text{ns}}=\frac{1}{4}\omega^2-\frac{2}{\alpha} f_n(\lambda)
=\frac{1}{4\alpha^2\lambda^2}-\frac{2}{\alpha} f_n(\lambda)  .
\end{eqnarray}   
Here, instead of parameter $\omega$, we use the parameter $\lambda=1/\omega\alpha$; functions $f_n(\lambda)$ dependent only on  variable $ \lambda$ are defined by 
\begin{eqnarray}\label{s56}
  f_n (\lambda)=u_{n-1,n-1}+ \frac{n-1}{2n-1} u_{n-1,n-2}+\frac{n}{2n-1}u_{n-1, n}.
\end{eqnarray} 

Plugging $\alpha=r_0/\kappa a_0^*$ into the formula, we have
\begin{eqnarray}\label{s57}
  E^{(0)}_{\text{ns}}=\frac{\kappa^2}{4 {r_0^*}^2\lambda^2}-\frac{2\kappa}{r_0^*} f_n(\lambda)  .
\end{eqnarray}   
where $r_0^*=r_0/a_0^*$ is a dimensionless screening length (in a unit of effective Bohr radius scaled by the dielectric constant). There is a free parameter $\lambda$ in energies \eqref{s57}, which we need to define. In principle, numerical exact solutions do not depend on this parameter, but approximate solutions are dependent on the choice of this parameter. Previous studies \cite{Hoangbook2015, PhysRevB2023} showed that choosing the free parameter around the solution of the equation  
\begin{eqnarray}\label{s58}
\frac{\partial E^{(0)}_n}{\partial \lambda}=0.
\end{eqnarray}
is optimal, leading to accurate energies within some low-lying corrections. We choose the parameter $\lambda$ from Eq.~\eqref{s58} and put it to expression \eqref{s57} to get energy in the zer0-order approximation. By numerical calculations with formulas \eqref{s51} and \eqref{s53}, we also obtain exciton energies in second- and third-order approximations. Material parameters are picked up from Table~\ref{tabS1}. Results in Tables \ref{tabS2}, \ref{tabS3}, \ref{tabS4}, and \ref{tabS5} for monolayers WSe$_2$, WS$_2$, MoSe$_2$, MoS$_2$ encapsulated in hBN slabs 
demonstrate extremely high precision, within 1.0 meV in second-order approximation and 0.5 meV in third-order approximation.
\begin{table}[htbp]
	\caption{\label{tabS2} Exciton energies (meV) calculated by regulated perturbation theory for monolayer WSe$_2$  encapsulated by hBN slabs with screening length $r_0=4.21$ nm, exciton reduced mass $\mu=0.19\, m_e$, dielectric constant $\kappa= 4.34$.}
	\begin{ruledtabular}
		\begin{tabular}{l r r r r r }
				&	$E^{(0)}$	&	$E^{(2)}$	& $E^{(3)}$ &	$E_{\text{num}}$ \cite{PhysRevB2023} \\
			\hline
			$1s$	&	$-164.98$	&	$-168.09$	&	$-168.55$	&	$-168.60$		\\
$2s$	&	$-39.46$	&	$-37.05$	&	$-38.52$	&	$-38.57$	\\
$3s$	&	$-16.94$	&	$-16.22$	&	$-16.40$   	&	$-16.56$		\\
$4s$	&	$-9.32$	&	$-9.03$	&	$-9.06$  	&	$-9.13$		\\
$5s$	&	$-5.87$	&	$-5.73$	&	$-5.74$   	&	$-5.77$		\\
		\end{tabular}
	\end{ruledtabular}
\end{table}

\begin{table}[htbp]
	\caption{\label{tabS3} Exciton energies (meV) calculated by regulated perturbation theory for monolayer WS$_2$  encapsulated by hBN slabs with screening length $r_0=3.76$ nm, exciton reduced mass $\mu=0.175\, m_e$, dielectric constant $\kappa= 4.16$.}
	\begin{ruledtabular}
		\begin{tabular}{l r r r r r }
				&	$E^{(0)}$	&	$E^{(2)}$	& $E^{(3)}$ &	$E_{\text{num}}$ \cite{PhysRevB2023} \\
			\hline
			$1s$	&	$-174.66$	&	$-177.92$	&	$-178.40$	&	$-178.62$		\\
$2s$	&	$-40.46$	&	$-38.25$	&	$-39.64$	&	$-39.73$	\\
$3s$	&	$-17.21$	&	$-16.55$	&	$-16.74$   	&	$-16.90$		\\
$4s$	&	$-9.43$	&	$-9.16$	&	$-9.20$   	&	$-9.28$		\\
$5s$	&	$-5.93$	&	$-5.79$	&	$-5.80 $  	&	$-5.85$		\\
		\end{tabular}
	\end{ruledtabular}
\end{table}

\begin{table}[htbp]
	\caption{\label{tabS4} Exciton energies (meV) calculated by regulated perturbation theory for monolayer MoSe$_2$  encapsulated by hBN slabs with screening length $r_0=3.90$ nm, exciton reduced mass $\mu=0.35\, m_e$, dielectric constant $\kappa= 4.40$.}
	\begin{ruledtabular}
		\begin{tabular}{l r r r r r }
				&	$E^{(0)}$	&	$E^{(2)}$	& $E^{(3)}$ &	$E_{\text{num}}$ \cite{PhysRevB2023} \\
			\hline
			$1s$	&	$-226.62$	&	$-231.01$	&	$-231.72$	&	$-231.96$		\\
$2s$	&	$-62.73$	&	$-57.05$	&	$-60.20$	&	$-60.63$	\\
$3s$	&	$-28.27$	&	$-26.53$	&	$-26.72$   	&	$-27.31$	\\
$4s$	&	$-15.88$	&	$-15.19$	&	$-15.14$   	&	$-15.41$		\\
$5s$	&	$-10.13$	&	$-9.79$	&	$-9.74$   	&		$-9.87$	\\
		\end{tabular}
	\end{ruledtabular}
\end{table}

\begin{table}[hb]
	\caption{\label{tabS5} Exciton energies (meV) calculated by regulated perturbation theory for monolayer MoS$_2$  encapsulated by hBN slabs with screening length $r_0=3.40$ nm, exciton reduced mass $\mu=0.275\, m_e$, dielectric constant $\kappa= 4.45$.}
	\begin{ruledtabular}
		\begin{tabular}{l r r r r r }
				&	$E^{(0)}$	&	$E^{(2)}$	& $E^{(3)}$ &	$E_{\text{num}}$ \cite{PhysRevB2023} \\
			\hline
$1s$	&	$-215.29$	&	$-219.38$	&	$-220.00$	&	$-220.18$		\\
$2s$	&	$-53.09$	&	$-49.52$	&	$-51.66$	&	$-51.81$	\\
$3s$	&	$-23.01$	&	$-21.94$	&	$-22.18$   	&	$-22.46$	\\
$4s$	&	$-12.71$	&	$-12.28$	&	$-12.27$   	&	$-12.44$		\\
$5s$	&	$-8.03$	&	$-7.82$	&	$-7.82$   	&		$-7.89$	\\
		\end{tabular}
	\end{ruledtabular}
\end{table}

\section{\label{SIII} Derivation of analytical expression for $s$-state exciton energies} 
{\bf{Exciton energy for the $1s$ state}}

For 1s state, Eq.~\eqref{s57} become 
\begin{eqnarray}\label{s59}
E^{(0)}_{1s}(r_0^*/\kappa)=\frac{\kappa^2}{4{r_0^*}^2}\frac{1}{\lambda^2}-\frac{2\kappa}{r_0^*}f_{1}(\lambda),
\end{eqnarray}
where 
\begin{eqnarray}\label{s60}
f_{1}(\lambda)=\frac{J(\lambda) +\lambda-1 }{\lambda^2+1}, \quad
J(\lambda)=\frac{ln{\left[\left(1+\sqrt{\lambda^{-2} +1}\right)\left(1+\sqrt{\lambda^{2} +1}\right)\right]}} {\sqrt{\lambda^2+1}}.
\end{eqnarray}
Here, $\lambda$ can be considered as a variational parameter and consequently is determined by the equation
\begin{eqnarray}\label{s61}
\frac{\partial E^{(0)}_{1s}}{\partial \lambda}=0.
\end{eqnarray}

Basically, solving Eqs.~\eqref{s61}, we can get an analytical formula for the $1s$-state energy. However, because of the complecity of function $J(\lambda)$ given by Eq.~\eqref{s60} the solving process is not trivial, requiring the Taylor expansion of this function. From another side, numerically solving Eq.~\eqref{s61} for the specific case of monolayers WSe$_2$, WS$_2$, MoSe$_2$, and MoS$_2$ encapsulated by hBN slabs with the material parameters given in Tab.~\ref{tabS1}), we get $\lambda_1=0.803, 0.863, 0.582,$ and $0.750$, corresponding to each material. These results suggest us to expand function $f_1(\lambda)$ around the value $\lambda_0=0.75$. This point is obtained by the formula $\frac{1}{2}(n+1/2)$ with $n=1$. The expansion is
\begin{eqnarray}\label{s62}
&f_1(\lambda)\simeq 2.454 -6.399\, \lambda +11.956\, \lambda^2-14.594\, \lambda^3
+11.079\, \lambda^4-4.757 \,\lambda^5+0.884 \,\lambda^6\quad.
\end{eqnarray}
Polynomial \eqref{s62} is truncated at $\lambda^{6}$ because its value for $\lambda$ in the range of $[0.5\rightarrow 0.9]$ is approximate to the function $f_1(\lambda)$ with an error less than $0.1\%$.

Plugging expansion \eqref{s62} for $f_1(\lambda)$ into Eq.~\eqref{s61}, we get an equation in the polynomial form for determining the parameter $\lambda$. The equation in term of $\delta \lambda = \lambda - 0.75$ is given as follows
\begin{eqnarray}\label{s63}
-\frac{\kappa}{r^*_0}+1.120+  2.326\,\delta \lambda +\cdots =0.
\end{eqnarray}
Approximately, we consider Eq.~\eqref{s63} in the first order of $\delta \lambda$ which leads to the solution for $\delta \lambda$ as
\begin{eqnarray}\label{s64}
\delta \lambda&=&0.430 \frac{\kappa}{r^*_0}-0.482.
\end{eqnarray}
Substituting the material data in Tab.~\ref{tabS1} into formula \eqref{s64}, we obtain the values of  $\lambda = 0.75+\delta \lambda$ as following: $0.804$ (0.803 WSe$_2$), $0.865$ (0.863 WS$_2$), $0.589$ (0.582 MoSe$_2$), and $0.748$ (0.750 MoS$_2$), which are consistent with numerically solving Eq.~\eqref{s61}  in direct (the values given in the parenthesis).

Using the calculated above formulae, we rewrite the formula \eqref{s59} for $1s$ state exciton energy in term of $\delta \lambda$ as
\begin{eqnarray}\label{s65}
E^{(0)}_{1s}&=&\frac{\kappa^2}{{r_0^*}^2}\frac{1}{(1.5+2\delta\lambda)^2}
                           -\frac{2\kappa}{r_0^*}\left( 0.757 -0.664\delta\lambda\right.\nonumber\\
&&\quad\left. +0.639 \delta\lambda^2-0.654 \delta\lambda^3+0.701\delta\lambda^4
 -0.778\delta\lambda^5 +0.884 \delta\lambda^6 \right).
\end{eqnarray}
From Eq.~\eqref{s64}, we obtain a formula $\kappa/r^*_0 = (\delta\lambda+0.482)/0.430$, which can be substituted 
into the second term of Eq.~\eqref{s65} to simplify this analytical exciton energy. As a result, formula \eqref{s65} becomes
\begin{eqnarray}\label{s66}
E^{(0)}_{1s}&=&-\frac{\kappa^2}{{r_0^*}^2}\frac{2.042}{(1.5+2\,\delta\lambda)^2}
                           \left( 1 -0.426\,\delta\lambda\right.\nonumber\\
&&\quad\left. +1.30\, \delta\lambda^2-2.956\, \delta\lambda^3+6.318\,\delta\lambda^4 -13.242\,\delta\lambda^5 +27.63\, \delta\lambda^6 \right).
\end{eqnarray}

Numerical calculations of formula \eqref{s66} give exciton energies with a precision of upto 5\% compared with the exact numerical solutions. So we will add to this formula the correction of second order of perturbation theory.  Especially, the correction of second-order approximation does not change the form of energy \eqref{s66}, but corrects only the coefficients in the equation. Indeed, putting the necessary calculated matrix elements into Eqs.~\eqref{s51} and then using the same Taylor expansion as above, we get
\begin{eqnarray}\label{s67}
E^{(2)}_{1s}&=&-\frac{\kappa^2}{{r_0^*}^2}\frac{2.08989}{(1.5+2\,\delta\lambda)^2}P_1(\delta\lambda),\\
P_1(\delta\lambda)&=& 1 -0.175\,\delta\lambda+0.742\, \delta\lambda^2-1.670\, \delta\lambda^3
+3.363\,\delta\lambda^4 -6.531\,\delta\lambda^5 +12.579\, \delta\lambda^6.\nonumber
\end{eqnarray}
For the second-order approximation energy \eqref{s67}, we  modify $\delta\lambda$ (Eq.~\eqref{s64}) slightly to get a general formula for all excited $s$-states as
\begin{eqnarray}\label{s68}
\delta \lambda_n&=&\left(0.470 \frac{\kappa}{r^*_0}-0.525\right) n.
\end{eqnarray}
This modification does not affect the final results much. 

Polynomial $P_1(\delta\lambda)$ in Eq.~\eqref{s67} is approximately equal to one because of the smallness of the derivation $\delta\lambda$  ($0.062$ WSe$_2$, $0.128$ WS$_2$, $-0.173$ MoSe$_2$, and $0.0003$ MoS$_2$). To evaluate this polynomial, we show in Tab.~\ref{tabS6} its values for different monolayer TMDCs. We can see that the polynomial has almost the value of 1.0 for MoS$_2$ and can be truncated at $\delta\lambda^0$. For monolayers WSe$_2$, this polynomial is equal to 0.99 and truncated at $\delta\lambda^1$. The same for monolayer WS$_2$, the polynomial has the value of 0.987 but is truncated at $\delta\lambda^3$. The worst result is for monolayer MoSe$_2$ where the polynomial has the value of 1.066 and is truncated at $\delta\lambda^5$.

\begin{table}[htbp]
	\caption{\label{tabS6} Values of the polynomial $P_1(\delta\lambda)$ with the derivation $\delta\lambda$ taken from the materal data for WSe$_2$, WS$_2$, MoSe$_2$, and MoS$_2$. $P_1^{(N)}(\delta\lambda)$ means the polynomial $P_1(\delta\lambda)$ truncated at the term $\sim {\delta\lambda}^{N}$. }
	\begin{ruledtabular}
		\begin{tabular}{l r r r r }
				&	WSe$_2$	&	WS$_2$  & MoSe$_2$ &	MoS$_2$ \\
			\hline
$\delta\lambda$ &	0.062 &	0.128 &	-0.173 &	0.0003 \\
\hline
$P_1^{(1)} (\delta\lambda)$	&	0.989   &	0.9775 &	1.0303	&	0.99994 \\
$P_1^{(2)} (\delta\lambda)$&	0.992	&	 0.9897 &	 1.0525  	&	0.99994	\\
$P_1^{(3)} (\delta\lambda)$ &	0.9917	&	0.9862  &	1.0611   	&	0.99994	\\
$P_1^{(4)} (\delta\lambda)$ &	0.9917	&	0.9871  &	   1.0641	&	0.99994	\\
$P_1^{(5)} (\delta\lambda)$ &	0.9917	&	0.9869  &	  1.0651 	&	0.99994	\\
$P_1^{(6)} (\delta\lambda)$ &	0.9917	&	0.9869  &	   1.0655	&		0.99994\\

		\end{tabular}
	\end{ruledtabular}
\end{table}

Now, plugging the explicit form of $\delta\lambda$ Eq.~\eqref{s68} with $n=1$ into Eq.~\eqref{s67} and using the truncation discussed above, we can rewrite the analytical energy \eqref{s67} in a general form for exciton energies as
\begin{eqnarray}\label{s69}
E^{(2)}_{1s}&=&-\frac{\text{Ry}^*\times P_1}{(0.650+0.311 \,r^*_0/\kappa)^2},
\end{eqnarray}
where $P_1$ is a polynomial whose value is around 1.0, as provided in Tab.~\ref{tabS6}. On the other hand, the energy spectra of the two-dimensional hydrogen atom have the form
\begin{eqnarray}\label{s70}
E_{n} &=&-\frac{\text{Ry}}{(n-1/2)^2}.
\end{eqnarray}
For the $1s$ state, $n=1$, so we expect the number $1-0.5=0.5$ instead of 0.569 in the denominator of Eq.~\eqref{s69}. To make the formula \eqref{s69} mimic the 2D hydrogen atom energy, we transform the main term in \eqref{s69} as follows:
\begin{eqnarray}
\frac{2.08989}{(1.5+2\,\delta\lambda)^2}
&=&\frac{2.08989}{(1.5+(2-x)\,\delta\lambda)^2}
\times\frac{1}{\left(1+\dfrac{x\delta\lambda}{1.5+(2-x)\,\delta\lambda}\right)^2}.\nonumber
\end{eqnarray}
Thus, the exciton energy now becomes
\begin{eqnarray}\label{s71}
E^{(2)}_{1s}=\frac{\kappa^2}{{r_0^*}^2}\frac{2.08989\,{\text{Ry}^*}}{(1.5+(2-x)\,\delta\lambda)^2}\times P_1^*(\delta\lambda)
\end{eqnarray}
with the polynomial $P_1$ modified as
\begin{eqnarray}\label{s72}
P_1^*(\delta\lambda)&=&\frac{P_1(\delta\lambda)}
           {\left(1+\dfrac{x\delta\lambda}{1.5+(2-x)\,\delta\lambda}\right)^2}.
\end{eqnarray}
Chosing the value $x=0.462$ and putting the explicit form of $\delta\lambda$ (Eq.~\eqref{s68} into the denominator of \eqref{s71}, we finally obtain
\begin{eqnarray}\label{s73}
E^{(2)}_{1s}&=&-\frac{\text{Ry}^*\times P^*_1}{(0.5+0.479 \,r^*_0/\kappa)^2}
\end{eqnarray}
with the polynomial
\begin{eqnarray}
P^*_1(\delta\lambda)&=& 1 -0.791\,\delta\lambda+1.766\, \delta\lambda^2-3.636\, \delta\lambda^3
+7.274\,\delta\lambda^4 -14.323\,\delta\lambda^5 +27.987\, \delta\lambda^6\nonumber
\end{eqnarray}
having values around 1.0 as shown in Table \ref{tabS7}.
\begin{table}[htbp]
	\caption{\label{tabS7} Values of the polynomial $P^*_1(\delta\lambda)$ with the derivation $\delta\lambda$ taken from the materal data for WSe$_2$, WS$_2$, MoSe$_2$, and MoS$_2$. $P_1^{(N)}(\delta\lambda)$ means the polynomial $P^*_1(\delta\lambda)$ truncated at the term $\sim {\delta\lambda}^{N}$. }
	\begin{ruledtabular}
		\begin{tabular}{l r r r r }
				&	WSe$_2$	&	WS$_2$  & MoSe$_2$ &	MoS$_2$ \\
			\hline
$\delta\lambda$ &	0.062 &	0.128 &	$-0.173$ &	0.0003 \\
\hline
$P_1^{(1)} (\delta\lambda)$	&	0.9511 &	0.8986 &	1.1368	&	0.9997 \\
$P_1^{(2)} (\delta\lambda)$&	0.9579	&	 0.9276 &	 1.1897  	&	0.9997	\\
$P_1^{(3)} (\delta\lambda)$ &	0.9570	&	0.9200  &	1.2085   	&	0.9997	\\
$P_1^{(4)} (\delta\lambda)$ &	0.9571	&	0.9219  &	   1.2150	&	0.9997	\\
$P_1^{(5)} (\delta\lambda)$ &	0.9571	&	0.9214  &	  1.2172 	&	0.9997	\\
$P_1^{(6)} (\delta\lambda)$ &	0.9571	&	0.9216  &	   1.2180	&		0.9997\\

		\end{tabular}
	\end{ruledtabular}
\end{table}

Table~\ref{tabS7} shows that polynomial $P^*_1$ truncated at $\sim \delta\lambda^4$ is accurate within a precision of 0.2\%. 
Table \ref{tabS8} gives exciton energies for the $1s$ state calculated by the analytical formula \eqref{s73} compared with the exact numerical solutions. The accuracy of the formula is very high, less than 1.0\%.

\begin{table}[htbp]
	\caption{\label{tabS8} Exciton energy of $1s$ state calculated by analytical formula \eqref{s73}. Comparison with the exact numerical solution \cite{PhysRevB2023} shows the high accuracy within errors less than 1\%.}
	\begin{ruledtabular}
		\begin{tabular}{l r r r r }
				&	WSe$_2$	&	WS$_2$  & MoSe$_2$ &	MoS$_2$ \\
			\hline
$r^*_0/\kappa $ &	0.803 &	0.719 &	1.333 &	0.893 \\
Ry$^*$ (meV)    &137.24& 137.59&245.97&188.94\\
\hline
Formula\eqref{s73} (meV)	&	$-168.03$ &	$-177.79$ &	$-230.79$	&	$-219.35$ \\
Exact num. \cite{PhysRevB2023} (meV)&	$-168.60$	&	 $-178.62$ &	 $-231.96$ &	$-220.18$ \\
Relative error (\%) &	0.34 &	 0.46 &	 0.50 	&	0.38\\

		\end{tabular}
	\end{ruledtabular}
\end{table}

{\bf{Analytical energies of $2s$ state}}

For 2s state, function $f_2(\lambda)$ has the form
\begin{eqnarray}\label{s74}
f_{2}(\lambda)&=&\frac{(1-3\lambda^2+\lambda^4)J(\lambda) }{(\lambda^2+1)^3}
+\frac{ -6+11\lambda +8\lambda^2-3\lambda^3-\lambda^4+\lambda^5}{3(\lambda^2+1)^3}.
\end{eqnarray}

By the same procedure as for $1s$-state, we obtain exciton energy for $2s$-state as
\begin{eqnarray}\label{s75}
E^{(2)}_{2s}&=&-\frac{\kappa^2}{{r_0^*}^2}\frac{1.325\,{\text{Ry}^*}}{(2.5+2\,\delta\lambda)^2}\times P_2(\delta\lambda),
\end{eqnarray}
where
\begin{eqnarray}\label{s76}
P_2(\delta\lambda)&=& 1 -0.0398\,\delta\lambda+0.0953\, \delta\lambda^2-0.119\, \delta\lambda^3+0.131\,\delta\lambda^4 -0.127\,\delta\lambda^5 +0.137\, \delta\lambda^6.
\end{eqnarray}
The equation $\partial E^0/\partial \lambda =0$ leads to $\delta\lambda =0.88 \kappa/{r_0^*}-1.04$. However, we modify it slightly following Eq.~\eqref{s68} to get
\begin{eqnarray}\label{s77}
\delta \lambda&=&0.94 \frac{\kappa}{r^*_0}-1.05.
\end{eqnarray}
Put it into Eq.~\eqref{s75} we obtain 
\begin{eqnarray}\label{s78}
E^{(2)}_{2s}&=&-\frac{\text{Ry}^*}{(1.633+0.348\,r^*_0/\kappa)^2}\times P_2(\delta\lambda),
\end{eqnarray}
where values of $P_2$ given in Table \ref{tabS9} are around 1.0.

\begin{table}[htbp]
	\caption{\label{tabS9} Values of the polynomial $P_2(\delta\lambda)$ with the derivation $\delta\lambda$ taken from the materal data for WSe$_2$, WS$_2$, MoSe$_2$, and MoS$_2$. $P_2^{(N)}(\delta\lambda)$ means the polynomial $P_2(\delta\lambda)$ truncated at the term $\sim {\delta\lambda}^{N}$. }
	\begin{ruledtabular}
		\begin{tabular}{l r r r r }
				&	WSe$_2$	&	WS$_2$  & MoSe$_2$ &	MoS$_2$ \\
			\hline
$\delta\lambda$ &	0.123 &	0.256 &	$-0.346$ &	0.0007 \\
\hline
$P_2^{(1)} (\delta\lambda)$	&	0.9951 &	0.9898 &	1.0138	&	0.99997 \\
$P_2^{(2)} (\delta\lambda)$&	0.9965	&	 0.9961 &	 1.0252  	&	0.99997	\\
$P_2^{(3)} (\delta\lambda)$ &	0.9963	&	0.9941  &	1.0301   	&	0.99997	\\
$P_2^{(4)} (\delta\lambda)$ &	0.9963	&	0.9946  &	   1.0320	&	0.99997	\\
$P_2^{(5)} (\delta\lambda)$ &	0.9963	&	0.9945  &	  1.0326 	&	0.99997	\\
$P_2^{(6)} (\delta\lambda)$ &	0.9963	&	0.9945  &	   1.0328	&		0.99997\\

		\end{tabular}
	\end{ruledtabular}
\end{table}

We use the following transformation 
\begin{eqnarray}
\frac{1.325}{(2.5+2\,\delta\lambda)^2}
&=&\frac{1.325}{(2.5+(2-x)\,\delta\lambda)^2}
\times\frac{1}{\left(1+\dfrac{x\delta\lambda}{2.5+(2-x)\,\delta\lambda}\right)^2}\nonumber
\end{eqnarray}
to modify the form of Eq.~\eqref{s75} as
\begin{eqnarray}\label{s79}
E^{(2)}_{2s}=-\frac{\kappa^2}{{r_0^*}^2}\frac{1.325\,{\text{Ry}^*}}{(2.5+(2-x)\,\delta\lambda)^2}\times P_2^*(\delta\lambda)
\end{eqnarray}
with the polynomial $P_2$ modified as
\begin{eqnarray}\label{s80}
P_2^*(\delta\lambda)&=&\frac{P_2(\delta\lambda)}
           {\left(1+\dfrac{x\delta\lambda}{2.5+(2-x)\,\delta\lambda}\right)^2}.
\end{eqnarray}
Chosing the value $x=0.163$ and putting the explicit form of $\delta\lambda$ (Eq.~\eqref{s77} into the denominator of \eqref{s79}, we finally obtain
\begin{eqnarray}\label{s81}
E^{(2)}_{2s}&=&-\frac{\text{Ry}^*}{(1.5+0.496 \,r^*_0/\kappa)^2}\times P^*_2
\end{eqnarray}
with
\begin{eqnarray}
P^*_2(\delta\lambda)&=& 1 -0.171\,\delta\lambda+0.209\, \delta\lambda^2-0.226\, \delta\lambda^3
+0.235\,\delta\lambda^4 -0.231\,\delta\lambda^5 +0.240\, \delta\lambda^6.\nonumber
\end{eqnarray}
Values of $P_2^*$ given in Table \ref{tabS10} are around 1.0.

\begin{table}[htbp]
	\caption{\label{tabS10} Values of the polynomial $P^*_2(\delta\lambda)$ with the derivation $\delta\lambda$ taken from the materal data for WSe$_2$, WS$_2$, MoSe$_2$, and MoS$_2$. $P_2^{(N)}(\delta\lambda)$ means the polynomial $P_2(\delta\lambda)$ truncated at the term $\sim {\delta\lambda}^{N}$. }
	\begin{ruledtabular}
		\begin{tabular}{l r r r r }
				&	WSe$_2$	&	WS$_2$  & MoSe$_2$ &	MoS$_2$ \\
			\hline
$\delta\lambda$ &	0.124 &	0.256 &	$-0.346$ &	0.0007 \\
\hline
$P_2^{(1)} (\delta\lambda)$	&	0.9789 &	0.9563 &	1.0590	&	0.99989 \\
$P_2^{(2)} (\delta\lambda)$&	0.9820	&	 0.9701 &	 1.0840  	&	0.99989	\\
$P_2^{(3)} (\delta\lambda)$ &	0.9817	&	0.9663  &	1.0934   	&	0.99989	\\
$P_2^{(4)} (\delta\lambda)$ &	0.9818	&	0.9673  &	   1.0967	&	0.99989	\\
$P_2^{(5)} (\delta\lambda)$ &	0.9818	&	0.9670  &	  1.0979 	&	0.99989	\\
$P_2^{(6)} (\delta\lambda)$ &	0.9818	&	0.9671  &	   1.0983	&		0.99989\\

		\end{tabular}
	\end{ruledtabular}
\end{table}

From Table~\ref{tabS10}, we can see that polynomial $P^*_2$ truncated at $\sim \delta\lambda^3$ is accurate within a precision of 0.2\%. 
Table \ref{tabS11} gives exciton energies for the $2s$ state calculated by analytical formula \eqref{s81} compared with the exact numerical solutions.

\begin{table}[htbp]
	\caption{\label{tabS11} Exciton energy of $2s$ state calculated by analytical formula \eqref{s81}. Comparison with the exact numerical solution \cite{PhysRevB2023} shows the high accuracy.}
	\begin{ruledtabular}
		\begin{tabular}{l r r r r }
				&	WSe$_2$	&	WS$_2$  & MoSe$_2$ &	MoS$_2$ \\
			\hline
$r^*_0/\kappa $ &	0.803 &	0.719 &	1.333 &	0.893 \\
Ry$^*$ (meV)    &137.24& 137.59&245.97&188.94\\
\hline
Formula\eqref{s81} (meV)	&	$-37.41$ &	$-38.59$ &	$-57.80$	&	$-50.03$ \\
Exact num. \cite{PhysRevB2023} (meV)&	$-38.57$	&	 $-39.73$ &	 $-60.63$ &	$-51.81$ \\
Relative error (\%) &	3.0 &	 2.9 &	 4.7 	&	3.4\\

		\end{tabular}
	\end{ruledtabular}
\end{table}

{\bf{Analytical energies of $3s$ state}}

For the $3s$ state, function $f_3 (\lambda) $ has the form
\begin{eqnarray}\label{s82}
f_{3}(\lambda)&=&\frac{q_3(\lambda)J(\lambda) +p_3(\lambda) }{(\lambda^2+1)^5}
\end{eqnarray}
with
\begin{eqnarray}
q_{3}(\lambda)&=&1-11\lambda^2+23.25\lambda^4-11\lambda^6+\lambda^8,\nonumber\\
p_3(\lambda)&=&-2.5+6.6\lambda+16.75\lambda^2-20.2\lambda^3-22.2\lambda^4 +18.75\lambda^5+5.6\lambda^6-1.5\lambda^7-0.2\lambda^8+0.2\lambda^9.\nonumber
\end{eqnarray}
After  the similar calculations, we get
\begin{eqnarray}\label{s83}
E^{(2)}_{3s}&=&-\frac{\kappa^2}{{r_0^*}^2}\frac{1.140\,\text{Ry}^*}{(3.5+2\,\delta\lambda)^2}\times P_3(\delta\lambda),
\end{eqnarray}
with 
\begin{eqnarray}
P_3(\delta\lambda)&=& 1 -0.0204\,\delta\lambda+0.0298\, \delta\lambda^2-0.0237\, \delta\lambda^3
+0.111\,\delta\lambda^4 -0.0478\,\delta\lambda^5 +0.00772\, \delta\lambda^6.\nonumber
\end{eqnarray}
The equation $\partial E^0/\partial \lambda =0$ leads to $\delta\lambda =1.38 \kappa/{r_0^*}-1.575$. However, we modify it slightly following Eq.~\eqref{s68} to get
\begin{eqnarray}\label{s84}
\delta \lambda&=&1.41 \frac{\kappa}{r^*_0}-1.575
\end{eqnarray}
Put it into Eq.~\eqref{s83}, we obtain 
\begin{eqnarray}\label{s85}
E^{(2)}_{3}&=&-\frac{\text{Ry}^*}{(2.642+0.327\,r^*_0/\kappa)^2}P_3(\delta\lambda),
\end{eqnarray}
Values of $P_3 (\delta\lambda)$ given in Table~\ref{tabS12} are around 1.0.
\begin{table}[htbp]
	\caption{\label{tabS12} Values of the polynomial $P_3(\delta\lambda)$ with the derivation $\delta\lambda$ taken from the materal data for WSe$_2$, WS$_2$, MoSe$_2$, and MoS$_2$. $P_1^{(N)}(\delta\lambda)$ means the polynomial $P_2(\delta\lambda)$ truncated at the term $\sim {\delta\lambda}^{N}$. }
	\begin{ruledtabular}
		\begin{tabular}{l r r r r }
				&	WSe$_2$	&	WS$_2$  & MoSe$_2$ &	MoS$_2$ \\
			\hline
$\delta\lambda$ &	0.185 &	0.364 &	$-0.519$ &	0.001 \\
$P_3^{(1)} (\delta\lambda)$	&	0.9962 &	0.9922 &	1.0106	&	0.99998 \\
$P_3^{(2)} (\delta\lambda)$&	0.9972	&	 0.9966 &	 1.0186  	&	0.99998	\\
$P_3^{(3)} (\delta\lambda)$ &	0.9971	&	0.9952  &	1.0219   	&	0.99998	\\
$P_3^{(4)} (\delta\lambda)$ &	0.9972	&	0.9976  &	   1.0299	&	0.99998	\\
$P_3^{(5)} (\delta\lambda)$ &	0.9972	&	0.9972  &	  1.0317 	&	0.99998	\\
$P_3^{(6)} (\delta\lambda)$ &	0.9972	&	0.9972  &	   1.0319	&		0.99998\\

		\end{tabular}
	\end{ruledtabular}
\end{table}

Further, we use the transformation
\begin{eqnarray}
\frac{1.140}{(3.5+2\,\delta\lambda)^2}
&=&\frac{1.140}{(3.5+(2-x)\,\delta\lambda)^2}
\times\frac{1}{\left(1+\dfrac{x\delta\lambda}{3.5+(2-x)\,\delta\lambda}\right)^2}.\nonumber
\end{eqnarray}
Thus, the exciton energy now become
\begin{eqnarray}\label{s86}
E^{(2)}_{3s}=\frac{\kappa^2}{{r_0^*}^2}\frac{1.140}{(3.5+(2-x)\,\delta\lambda)^2}\times P_3^*(\delta\lambda)
\end{eqnarray}
with the polynomial $P_3$ modified as
\begin{eqnarray}\label{s87}
P_3^*(\delta\lambda)&=&\frac{P_3(\delta\lambda)}
           {\left(1+\dfrac{x\delta\lambda}{3.5+(2-x)\,\delta\lambda}\right)^2}.
\end{eqnarray}

Chosing the value $x=0.107$ and putting the explicit form of $\delta\lambda$ (Eq.~\eqref{s84} into the denominator of \eqref{s86}, we finally obtain
\begin{eqnarray}\label{s88}
E^{(2)}_{3s}&=&-\frac{\text{Ry}^*}{(2.5+0.486 \,r^*_0/\kappa)^2}\times P^*_3,
\end{eqnarray}
where
\begin{eqnarray}\label{s89}
P^*_3(\delta\lambda)&=& 1 -0.0816\,\delta\lambda+0.0670\, \delta\lambda^2-0.0473\, \delta\lambda^3+0.126\,\delta\lambda^4 -0.0635\,\delta\lambda^5 +0.0199\, \delta\lambda^6.
\end{eqnarray}

Values given in Table~\ref{tabS13} show that the function $P_3^*(\delta\lambda)$ is almost 1.0.

\begin{table}[htbp]
	\caption{\label{tabS13} Values of the polynomial $P^*_3(\delta\lambda)$ with the derivation $\delta\lambda$ taken from the materal data for WSe$_2$, WS$_2$, MoSe$_2$, and MoS$_2$. $P_3^{(N)}(\delta\lambda)$ means the polynomial $P^*_3(\delta\lambda)$ truncated at the term $\sim {\delta\lambda}^{N}$. }
	\begin{ruledtabular}
		\begin{tabular}{l r r r r }
				&	WSe$_2$	&	WS$_2$  & MoSe$_2$ &	MoS$_2$ \\
			\hline
$\delta\lambda$ &	0.185 &	0.384 &	$-0.519$ &	0.001 \\
\hline
${P_3^*}^{(1)} (\delta\lambda)$	&	0.9849 &	0.9686 &	1.0424	&	0.99992 \\
${P_3^*}^{(2)} (\delta\lambda)$&	0.9872	&	 0.9785 &	 1.0604  	&	0.99992	\\
${P_3^*}^{(3)} (\delta\lambda)$ &	0.9869	&	0.9758  &	1.0670   	&	0.99992	\\
${P_3^*}^{(4)} (\delta\lambda)$ &	0.9870	&	0.9786  &	   1.0761	&	0.99992	\\
${P_3^*}^{(5)} (\delta\lambda)$ &	0.9870	&	0.9781  &	  1.0785	&	0.99992	\\
${P_3^*}^{(6)} (\delta\lambda)$ &	0.9870	&	0.9781  &	   1.0789	&		0.99992\\

		\end{tabular}
	\end{ruledtabular}
\end{table}

\begin{table}[htbp]
	\caption{\label{tabS14} Exciton energy of $3s$ state calculated by analytical formula \eqref{s88} with polynomial $P^*_3$ by Eq.~\eqref{s89}. Comparison with the exact numerical solution \cite{PhysRevB2023} shows the high accuracy.}
	\begin{ruledtabular}
		\begin{tabular}{l r r r r }
				&	WSe$_2$	&	WS$_2$  & MoSe$_2$ &	MoS$_2$ \\
			\hline
$r^*_0/\kappa $ &	0.803 &	0.719 &	1.333 &	0.893 \\
Ry$^*$ (meV)    &137.24& 137.59&245.97&188.94\\
\hline
Formula\eqref{s88} (meV)	&	$-16.22$ &	$-16.58$ &	$-26.77$	&	$-21.94$ \\
Exact num. \cite{PhysRevB2023} (meV)&	$-16.56$	&	 $-16.90$ &	 $-27.31$ &	$-22.46$ \\
Relative error (\%) &	2.1 &	 1.9 &	 2.0 	&	2.3\\

		\end{tabular}
	\end{ruledtabular}
\end{table}

{\bf{Analytical energies of $4s$ state}}

For the $4s$ state, function $f_4(\lambda)$ has the form
\begin{eqnarray}\label{s90}
f_{4}(\lambda)&=&\frac{21q_{4}(\lambda) J(\lambda) + p_{4}(\lambda)}{21(\lambda^2+1)^7},
\end{eqnarray}
where
\begin{eqnarray}
q_4(\lambda)&=&1-24\lambda^2+131.25\lambda^4-223.75\lambda^6+131.25\lambda^8-24\lambda^{10}+\lambda^{12}\nonumber\\
p_4(\lambda)&=&-59.5+201\lambda+964.25\lambda^2-1696\lambda^3-3774.75\lambda^4
+4625.75\lambda^5+4520.75\lambda^6\nonumber\\
&&-3652.75\lambda^7-1759\lambda^8+1027.25\lambda^9+180\lambda^{10}-38.5\lambda^{11}-3\lambda^{12}+3\lambda^{13}.\nonumber
\end{eqnarray}
We obtain the exciton energy
\begin{eqnarray}\label{s91}
E^{(2)}_{4s}&=&-\frac{\kappa^2}{{r_0^*}^2}\frac{1.047}{(4.5+2\,\delta\lambda)^2}\times P_4(\delta\lambda),
\end{eqnarray}
where
\begin{eqnarray}
P_4(\delta\lambda)&=& 1 -0.0106\,\delta\lambda+0.00858\, \delta\lambda^2-0.00749\, \delta\lambda^3+0.00017\,\delta\lambda^4 -0.0118\,\delta\lambda^5 +0.00086\, \delta\lambda^6.\nonumber
\end{eqnarray}

Solution of equation $\partial E^{0}/\partial \lambda=0$ leads to  $\delta \lambda=1.88 {\kappa}/{r^*_0}-2.09$; however, we slightly modify it for a general formulation to get
\begin{eqnarray}\label{s92}
\delta \lambda&=&1.88 \frac{\kappa}{r^*_0}-2.1
\end{eqnarray}
We ontain the exciton energy in the second-order approximation as
\begin{eqnarray}\label{s93}
E^{(2)}_{4s}&=&-\frac{1}{(3.66+0.305\,r^*_0/\kappa)^2}P_4(\delta\lambda),
\end{eqnarray}
where the values of $P_4$ given in Table~\ref{tabS15} are around 1.0.
\begin{table}[htbp]
	\caption{\label{tabS15} Values of the polynomial $P_4(\delta\lambda)$ with the derivation $\delta\lambda$ taken from the materal data for WSe$_2$, WS$_2$, MoSe$_2$, and MoS$_2$. $P_1^{(N)}(\delta\lambda)$ means the polynomial $P_2(\delta\lambda)$ truncated at the term $\sim {\delta\lambda}^{N}$. }
	\begin{ruledtabular}
		\begin{tabular}{l r r r r }
				&	WSe$_2$	&	WS$_2$  & MoSe$_2$ &	MoS$_2$ \\
			\hline
$\delta\lambda$ &	0.247 &	0.512 &	$-0.692$ &	0.013 \\
$P_4^{(1)} (\delta\lambda)$	&	0.9974   &	0.9946 &	1.0074	&	0.9999 \\
$P_4^{(2)} (\delta\lambda)$&	0.9979	&	 0.9968 &	 1.0115  	&		0.9999\\
$P_4^{(3)} (\delta\lambda)$ &	0.9978	&	0.9958  &	1.0139   	&	0.9999	\\
$P_4^{(4)} (\delta\lambda)$ &	0.9978	&	0.9958  &	   1.0140	&	0.9999	\\
$P_4^{(5)} (\delta\lambda)$ &	0.9978	&	0.9954  &	  1.0159 	&	0.9999	\\
$P_4^{(6)} (\delta\lambda)$ &	0.9978	&	0.9954  &	   1.0159	&		0.9999\\

		\end{tabular}
	\end{ruledtabular}
\end{table}

We make a transformation
\begin{eqnarray}
\frac{1.047}{(4.5+2\,\delta\lambda)^2}
&=&\frac{1.047}{(4.5+(2-x)\,\delta\lambda)^2}\times\frac{1}{\left(1+\dfrac{x\delta\lambda}{4.5+(2-x)\,\delta\lambda}\right)^2}\nonumber
\end{eqnarray}
to get exciton energy
\begin{eqnarray}\label{s94}
E^{(2)}_{4s}=-\frac{\kappa^2}{{r_0^*}^2}\frac{1.047\,\text{Ry}^*}{(4.5+(2-x)\,\delta\lambda)^2}\times P_4^*(\delta\lambda)
\end{eqnarray}
with the polynomial $P_4$ modified as
\begin{eqnarray}\label{s95}
P_4^*(\delta\lambda)&=&\frac{P_4(\delta\lambda)}
           {\left(1+\dfrac{x\delta\lambda}{4.5+(2-x)\,\delta\lambda}\right)^2}.
\end{eqnarray}

Chosing the value $x=0.095$ and putting the explicit form of $\delta\lambda$ (Eq.~\eqref{s92} into the denominator of \eqref{s94}, we finally obtain
\begin{eqnarray}\label{s96}
E^{(2)}_{4s}&=&-\frac{\text{Ry}^*}{(3.5+0.488 \,r^*_0/\kappa)^2}\times P^*_4
\end{eqnarray}
with
\begin{eqnarray}\label{s97}
P^*_4(\delta\lambda)&=& 1 -0.0528\,\delta\lambda+0.0282\, \delta\lambda^2-0.0168\, \delta\lambda^3+0.0047\,\delta\lambda^4 -0.0139\,\delta\lambda^5 +0.0023\, \delta\lambda^6.
\end{eqnarray}
Values of $P_4^*$ given in Table~\ref{tabS16} are almost 1.0.
\begin{table}[htbp]
	\caption{\label{tabS16} Values of the polynomial $P^*_4(\delta\lambda)$ with the derivation $\delta\lambda$ taken from the materal data for WSe$_2$, WS$_2$, MoSe$_2$, and MoS$_2$. $P_1^{(N)}(\delta\lambda)$ means the polynomial $P^*_4(\delta\lambda)$ truncated at the term $\sim {\delta\lambda}^{N}$. }
	\begin{ruledtabular}
		\begin{tabular}{l r r r r }
				&	WSe$_2$	&	WS$_2$  & MoSe$_2$ &	MoS$_2$ \\
			\hline
$\delta\lambda$ &	0.247 &	0.512 &	$-0.692$ &	0.0013 \\
\hline
${P_4^*}^{(1)} (\delta\lambda)$	&	0.9870   &	0.9729 &	1.0365	&	0.99993 \\
${P_4^*}^{(2)} (\delta\lambda)$&	0.9887	  &	 0.9803     &	 1.0501  	&		0.99993\\
${P_4^*}^{(3)} (\delta\lambda)$ &	0.9884	&	0.9781  &	1.0556   	&	0.99993	\\
${P_4^*}^{(4)} (\delta\lambda)$ &	0.9884	&	0.9784  &	   1.0566	&	0.99993	\\
${P_4^*}^{(5)} (\delta\lambda)$ &	0.9884	&	0.9779  &	  1.0589 	&	0.99993	\\
${P_4^*}^{(6)} (\delta\lambda)$ &	0.9884	&	0.9779  &	   1.0591	&		0.99993\\

		\end{tabular}
	\end{ruledtabular}
\end{table}

\begin{table}[htbp]
	\caption{\label{tabS17} Exciton energy of $4s$ state calculated by analytical formula \eqref{s96} with polynomial $P^*_4$ by Eq.~\eqref{s97}. Comparison with the exact numerical solution \cite{PhysRevB2023} shows the high accuracy.}
	\begin{ruledtabular}
		\begin{tabular}{l r r r r }
				&	WSe$_2$	&	WS$_2$  & MoSe$_2$ &	MoS$_2$ \\
			\hline
$r^*_0/\kappa $ &	0.803 &	0.719 &	1.333 &	0.893 \\
Ry$^*$ (meV)    &137.24& 137.59&245.97&188.94\\
\hline
Formula\eqref{s96} (meV)	&	$-8.96$ &	$-9.07$ &	$-15.12$	&	$-12.20$ \\
Exact num. \cite{PhysRevB2023} (meV)&	$-9.13$	&	 $-9.28$ &	 $-15.41$ &	$-12.44$ \\
Relative error (\%) &	1.9 &	 2.2 &	 2.0 	&	1.9\\

		\end{tabular}
	\end{ruledtabular}
\end{table}

{\bf{Analytical energies of $5s$ state}}

For the $5s$ state, we have
\begin{eqnarray}\label{s98}
f_{5}(\lambda)&=&\frac{9q_{5}(\lambda) J(\lambda) + p_{5}(\lambda)}{9(\lambda^2+1)^9},
\end{eqnarray}
with
\begin{eqnarray}
q_5(\lambda)&=&1-42\lambda^2+436.75\lambda^4-1612.75\lambda^6+2462.27\lambda^8-1612.75\lambda^{10}+436.75\lambda^{12}-42\lambda^{14}+\lambda^{16}\nonumber\\
p_5(\lambda)&=&-2.75+113\lambda+829.88\lambda^2-1854\lambda^3-6561.66\lambda^4+10369.25\lambda^5+18599.11\lambda^6
-21400.75\lambda^7-21526.75\lambda^8\nonumber\\
&&+18725.11\lambda^9+10285.25\lambda^{10}-6477.66\lambda^{11}-1890\lambda^{12}+865.88\lambda^{13}+104\lambda^{14}-18.75\lambda^{15}-\lambda^{16}+\lambda^{17}.\nonumber
\end{eqnarray}

Solution of equation $\partial E^{0}/\partial \lambda=0$ leads to  $\delta \lambda=2.37\, {\kappa}/{r^*_0}-2.587$; however, we slightly modify it for a general formulation to get
\begin{eqnarray}\label{s99}
\delta \lambda&=&2.35 \frac{\kappa}{r^*_0}-2.625.
\end{eqnarray}

We have exciton energy for $5s$ state as
\begin{eqnarray}\label{s102}
E^{(2)}_{5s}&=&-\frac{\text{Ry}^*}{(4.723+0.251\,r^*_0/\kappa)^2}\times P_5(\delta\lambda),
\end{eqnarray}
where
\begin{eqnarray}\label{s101}
P_5(\delta\lambda)&=& 1 -0.0034\,\delta\lambda+0.000005\, \delta\lambda^2-0.0046\, \delta\lambda^3+0.00027\,\delta\lambda^4 -0.000006\,\delta\lambda^5 +0.00007\, \delta\lambda^6.
\end{eqnarray}
has the values around 1,0 shown in Table \ref{tabS18}
\begin{table}[htbp]
	\caption{\label{tabS18} Values of the polynomial $P_5(\delta\lambda)$ with the derivation $\delta\lambda$ taken from the materal data for WSe$_2$, WS$_2$, MoSe$_2$, and MoS$_2$. $P_5^{(N)}(\delta\lambda)$ means the polynomial $P_5(\delta\lambda)$ truncated at the term $\sim {\delta\lambda}^{N}$. }
	\begin{ruledtabular}
		\begin{tabular}{l r r r r }
				&	WSe$_2$	&	WS$_2$  & MoSe$_2$ &	MoS$_2$ \\
			\hline
$\delta\lambda$ &	0.309 &	0.641 &	$-0.865$ &	0.0017 \\
$P_5^{(1)} (\delta\lambda)$	&	0.9990   &	0.9978 &	1.0029	&	0.99999 \\
$P_5^{(2)} (\delta\lambda)$&	0.9990	&	 0.9979 &	 1.0029  	&		0.99999\\
$P_5^{(3)} (\delta\lambda)$ &	0.9988	&	0.9966  &	1.0059  	&	0.99999	\\
$P_5^{(4)} (\delta\lambda)$ &	0.9988	&	0.9967  &	   1.0060	&	0.99999	\\
$P_5^{(5)} (\delta\lambda)$ &	0.9988	&	0.9967  &	  1.0060 	&	0.99999	\\
$P_5^{(6)} (\delta\lambda)$ &	0.9988	&	0.9967  &	   1.0060	&		0.99999\\

		\end{tabular}
	\end{ruledtabular}
\end{table}

After the transformation
\begin{eqnarray}
\frac{0.9904}{(5.5+2\,\delta\lambda)^2}
&=&\frac{0.9904}{(5.5+(2-x)\,\delta\lambda)^2}\times\frac{1}{\left(1+\dfrac{x\delta\lambda}{5.5+(2-x)\,\delta\lambda}\right)^2},
\end{eqnarray}
the exciton energy becomes
\begin{eqnarray}\label{s103}
E^{(2)}_5=\frac{\kappa^2}{{r_0^*}^2}\frac{0.9904\,\text{Ry}^*}{(5.5+(2-x)\,\delta\lambda)^2}\times P_5^*(\delta\lambda)
\end{eqnarray}
with the polynomial $P_5$ modified as
\begin{eqnarray}\label{s104}
P_5^*(\delta\lambda)&=&\frac{P_5(\delta\lambda)}
           {\left(1+\dfrac{x\delta\lambda}{5.5+(2-x)\,\delta\lambda}\right)^2}.
\end{eqnarray}

Chosing the value $x=0.0944$ and putting the explicit form of $\delta\lambda$ (Eq.~\eqref{s99} into the denominator of \eqref{s103}, we finally obtain
\begin{eqnarray}\label{s105}
E^{(2)}_{5s}&=&-\frac{\text{Ry}^*}{(4.5+0.5001 \,r^*_0/\kappa)^2}\times P^*_5,
\end{eqnarray}
where
\begin{eqnarray}\label{s106}
P^*_5(\delta\lambda)&=& 1 -0.0377\,\delta\lambda+0.0129\, \delta\lambda^2-0.0094\, \delta\lambda^3+0.0022\,\delta\lambda^4 -0.0007\,\delta\lambda^5 +0.00034\, \delta\lambda^6
\end{eqnarray}
has values around 1.0 shown in Table~\ref{tabS19}

\begin{table}[htbp]
	\caption{\label{tabS19} Values of the polynomial $P^*_5(\delta\lambda)$ with the derivation $\delta\lambda$ taken from the materal data for WSe$_2$, WS$_2$, MoSe$_2$, and MoS$_2$. $P_5^{(N)}(\delta\lambda)$ means the polynomial $P^*_5(\delta\lambda)$ truncated at the term $\sim {\delta\lambda}^{N}$. }
	\begin{ruledtabular}
		\begin{tabular}{l r r r r }
				&	WSe$_2$	&	WS$_2$  & MoSe$_2$ &	MoS$_2$ \\
			\hline
$\delta\lambda$ &	0.309 &	0.641 &	$-0.865$ &	0.0017 \\
${P_5^*}^{(1)} (\delta\lambda)$	&	0.9884  &	0.9759 &	1.0326	&	0.99994 \\
${P_5^*}^{(2)} (\delta\lambda)$&	0.9896	&	 0.9812 &	 1.0422  	&		0.99994\\
${P_5^*}^{(3)} (\delta\lambda)$ &	0.9893	&	0.9787 &	1.0483  	&	0.99994	\\
${P_5^*}^{(4)} (\delta\lambda)$ &	0.9893	&	0.9791 &	   1.0495	&	0.99994	\\
${P_5^*}^{(5)} (\delta\lambda)$ &	0.9893	&	0.9790  &	  1.0485 	&	0.99994	\\
${P_5^*}^{(6)} (\delta\lambda)$ &	0.9893	&	0.9790  &	   1.0500	&		0.99994\\

		\end{tabular}
	\end{ruledtabular}
\end{table}

\begin{table}[htbp]
	\caption{\label{tabS20} Exciton energy of $5s$ state calculated by analytical formula \eqref{s105} with polynomial $P^*_4$ by Eq.~\eqref{s106}. Comparison with the exact numerical solution \cite{PhysRevB2023} shows the high accuracy within errors less than 1\%.}
	\begin{ruledtabular}
		\begin{tabular}{l r r r r }
				&	WSe$_2$	&	WS$_2$  & MoSe$_2$ &	MoS$_2$ \\
			\hline
$r^*_0/\kappa $ &	0.801 &	0.719 &	1.334 &	0.894 \\
Ry$^*$ (meV)    &137.24& 137.59&245.97&188.94\\
\hline
Formula\eqref{s105} (meV)	&	$-5.65$ &	$-5.70$ &	$-9.67$	&	$-7.72$ \\
Exact num. \cite{PhysRevB2023} (meV)&	$-5.77$	&	 $-5.85$ &	 $-9.87$ &	$-7.89$ \\
Relative error (\%) &	2.1 &	 2.6 &	 2.0 	&	2.1\\

		\end{tabular}
	\end{ruledtabular}
\end{table}

From Eqs.~\eqref{s73}, \eqref{s81}, \eqref{s88}, \eqref{s96}, and \eqref{s105}  for analytical energies of $1s$, $2s$, $3s$, $4s$, and $5s$, we combine them into a general formula
\begin{eqnarray}\label{s107}
E_{ns}&=&-\frac{\text{Ry}^*}{(n-0.5+0.479 \,r^*_0/\kappa)^2}\times P^*_n.
\end{eqnarray}
Here, from different values in the denominators as 0.479, 0.496, 0.486, 0.488, 0.500 we choose one value 0.479 because this value makes the formula best fit with the exact numerical solutions. From different polynomials $P_1^*$, $P_2^*$, $P_3^*$, $P_4^*$, and $P_5^*$, we can unify them into one function as follows.  First, we estimate the polynomials by functions as 
\begin{eqnarray}\label{s108}
P^*_1(\delta\lambda)&=& 1 -0.791\,\delta\lambda+1.766\, \delta\lambda^2-3.636\, \delta\lambda^3+7.274\,\delta\lambda^4 \sim\frac{\exp{(0.914\, \delta\lambda)}}{1+1.705 \,\delta\lambda},
\end{eqnarray}
\begin{eqnarray}\label{s109}
P^*_2(\delta\lambda)&=& 1 -0.171\,\delta\lambda+0.209\, \delta\lambda^2-0.226\, \delta\lambda^3+0.235\,\delta\lambda^4 \sim\frac{\exp{(0.4545\, \delta\lambda)}}{1+0.625 \,\delta\lambda},
\end{eqnarray}
\begin{eqnarray}\label{s110}
P^*_3(\delta\lambda)&=& 1 -0.0816\,\delta\lambda+0.0670\, \delta\lambda^2-0.0473\, \delta\lambda^3+0.126\,\delta\lambda^4 \sim\frac{\exp{(0.275\, \delta\lambda)}}{1+0.3566 \,\delta\lambda},
\end{eqnarray}
\begin{eqnarray}\label{s111}
P^*_4(\delta\lambda)&=& 1 -0.0528\,\delta\lambda+0.0282\, \delta\lambda^2-0.0168\, \delta\lambda^3\sim\frac{\exp{(0.179\, \delta\lambda)}}{1+0.2318 \,\delta\lambda},
\end{eqnarray}
\begin{eqnarray}\label{s112}
P^*_5(\delta\lambda)&=& 1 -0.0377\,\delta\lambda+0.0129\, \delta\lambda^2-0.0094\, \delta\lambda^3\sim\frac{\exp{(0.1183\, \delta\lambda)}}{1+0.156 \,\delta\lambda}.
\end{eqnarray}

Then, we rewrite them in one general function as
\begin{eqnarray}\label{s113}
P^*_n(r_0^*/\kappa)&=& \frac{\exp{(a_n\, \delta)}}{1+b_n \,\delta},
\end{eqnarray}
where
\begin{eqnarray}\label{s114}
\delta=\frac{\delta\lambda}{n}=0.47\kappa/r_0^*-0.525,\quad
a_n=\frac{4.3}{n}-\frac{6.6}{n^2}+\frac{3.2}{n^3}, \quad b_n=\frac{5.3}{n}-\frac{7.6}{n^2}+\frac{4.0}{n^3}.
\end{eqnarray}

\begin{table}[htbp]
	\caption{\label{tabS21} Exciton energies of $ns$ states calculated by the analytical formula \eqref{s107} compared with the numerical exact solutions in Ref.~\cite{PhysRevB2023} for monolayer TMDCs encapsulated by hBN.}
	\begin{ruledtabular}
		\begin{tabular}{l r r r r }
				&	WSe$_2$	&	WS$_2$  & MoSe$_2$ &	MoS$_2$ \\
			\hline
$r_0^*/\kappa $ &	0.803 &	0.719 &	1.333 &	0.893 \\
\hline
$E_{\text{1s}}$ (meV)	&	$-167.96$ &	$-177.80$ &	$-229.88$	&	$-219.34$ \\
Ref.~\cite{PhysRevB2023}  &	$-168.60$	&	 $-178.62$ &	 $-231.96$ &	$-220.18$ \\
\hline
$E_{\text{2s}}$ (meV)	&	$-37.94$ &	$-39.12$ &	$-58.71$	&	$-50.83$ \\
Ref.~\cite{PhysRevB2023}  &	$-38.57$	&	 $-39.73$ &	 $-60.63$ &	$-51.81$ \\
\hline
$E_{\text{3s}}$ (meV)	&	$-16.28$ &	$-16.61$ &	$-26.60$	&	$-22.04$ \\
Ref.~\cite{PhysRevB2023}  &	$-16.56$	&	 $-16.90$ &	$-27.31$ &	$-22.46$ \\
\hline
$E_{\text{4s}}$ (meV)	&	$-9.00$ &	$-9.13$ &	$-15.07$	&	$-12.25$ \\
Ref.~\cite{PhysRevB2023}  &	$-9.13$	&	 $-9.28$ &	$ -15.41$ &	$-12.44$ \\
\hline
$E_{\text{5s}}$ (meV)	&	$-5.70$ &	$-5.77$ &	$-9.68$	&	$-7.78$ \\
Ref.~\cite{PhysRevB2023}  &	$-5.77$	&	 $-5.85$ &	$ -9.87$ &	$-7.89$ \\

		\end{tabular}
	\end{ruledtabular}
\end{table}

We can also compare analytical energies \eqref{s107} with the formula suggested by Molas {\it{et al.}} \cite{Molas2019} as
\begin{eqnarray}\label{s107}
E_{ns}^{\text{Molas}}&=&-\frac{\text{Ry}^*}{(n+\delta)^2}.
\end{eqnarray}
Our formula is similar to the one of Molas with $\delta=-0.5+0.479*r_0^*/\kappa$ and $\text{Ry}^*(\text{modify})={\text{Ry}}^*\times P_n$. For monolayer WSe$_2$ in bHN, Molas's formula is $E_{ns}^{\text{Molas}}=134 {\text{meV}}/(n-0.099)^2$ while our formula is rewritten as $E_{ns}=134.5 {\text{meV}} /(1-0.11)^2$. Notably, in our formula, the effect of screening the charge is presented via the correction $P_n^*$. 

\section{Experimental validation}

We use experimental data of Refs. \cite{PRL2018,NAT2019,Liu2019} to test analytical formula \eqref{s107}. Some data are provided numerically in these works, and we extract other data from the figures of the works. Theoretically, we calculate exciton energies for monolayer Wse$_2$, WS$_2$, MoSe$_2$, and MoS$_2$ by formula \eqref{s107} with material parameters picked up from Refs.~\cite{NAT2019} and \cite{PhysRevB2023}. 
For MoTe$_2$, Ref.~\cite{NAT2019} give only two energy values: $E_b=177\pm 3$ meV and transition energy $2s-1s$ 124 meV. Material parameters given in Ref.~\cite{NAT2019}: $\mu=0.36 \pm m_e$, $r_0=6.4 \pm 0.3$ nm are in a wide interval of uncertainty while $\kappa=4.4$ is picked up from other works. Choosing the material parameters so that our theory best fits with the experimental data, we have $\mu=0.37 m_e$, $\kappa=4.3$, and $r_0=6.1$ nm. We show the experimental data and theoretical exciton energies in Figs.~\ref{fig1}, \ref{fig2}, and \ref{fig3}, which demonstrate the high consistency of our theory with the experiments. Experimental data of Liu {\it{et al.}} \cite{Liu2019} are not shown in the figures because they are almost identified with the ones of Ref.~\cite{PRL2018}. Numerically, they match our theory. Indeed, for monolayer WSe$_2$ , $E_{1s}=-172$ meV, $E_{2s}= -41$ meV, $E_{3s}=-20$ meV, $E_{4s}=-11$ meV. They lead to the transition energies $2s-1s$ 131 meV,  $3s-2s$ 21 meV,  $4s-2s$ 30 meV. Our theory  give $-168.38$, $-38.34$,   $-16.32$, and $- 8.97$  meV for energies, and  $2s-1s$ 130.04 meV,  $3s-2s$ 22.02 meV,  $4s-2s$ 29.37 meV for transition energies. 

\begin{figure}[htbp]
\begin{center}
\includegraphics[width=0.4 \columnwidth]{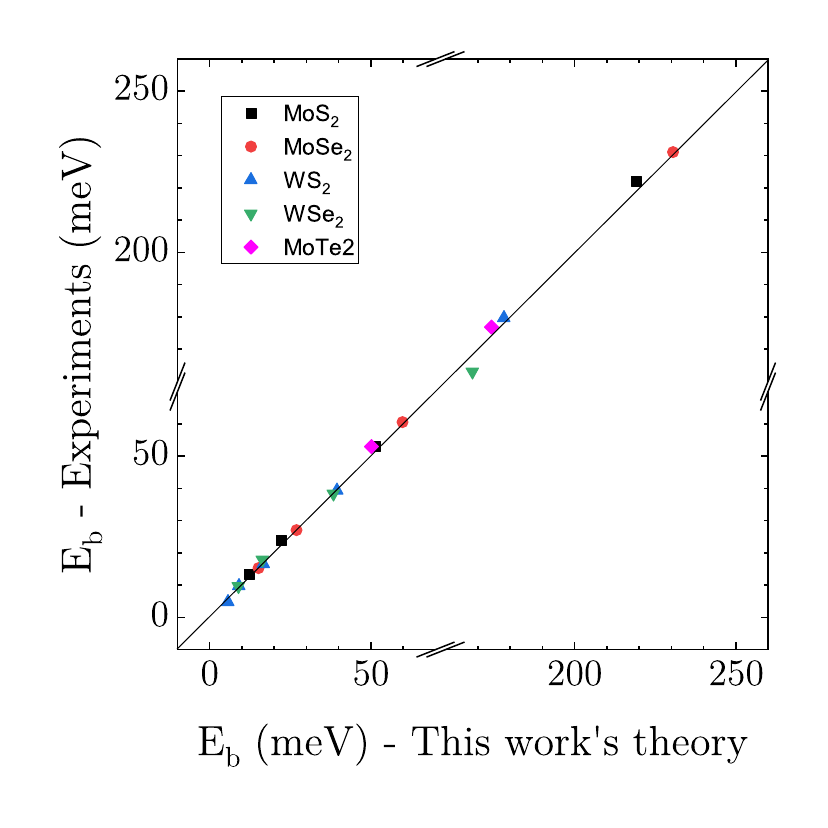}
\caption{Exciton energies for the $ns$ states in monolayers WSe$_2$, WS$_2$, MoSe$_2$, MoS$_2$, and MoTe$_2$ calculated by formula \eqref{s107} (horizontal axis) compared with the experimental data \cite{NAT2019, PRL2018} (vertical axis). Material parameters used for calculations are provided in Table  \ref{tabS1}. }
\label{fig1}
\end{center}
\end{figure} 

\begin{figure}[htbp]
\begin{center}
\includegraphics[width=0.7 \columnwidth]{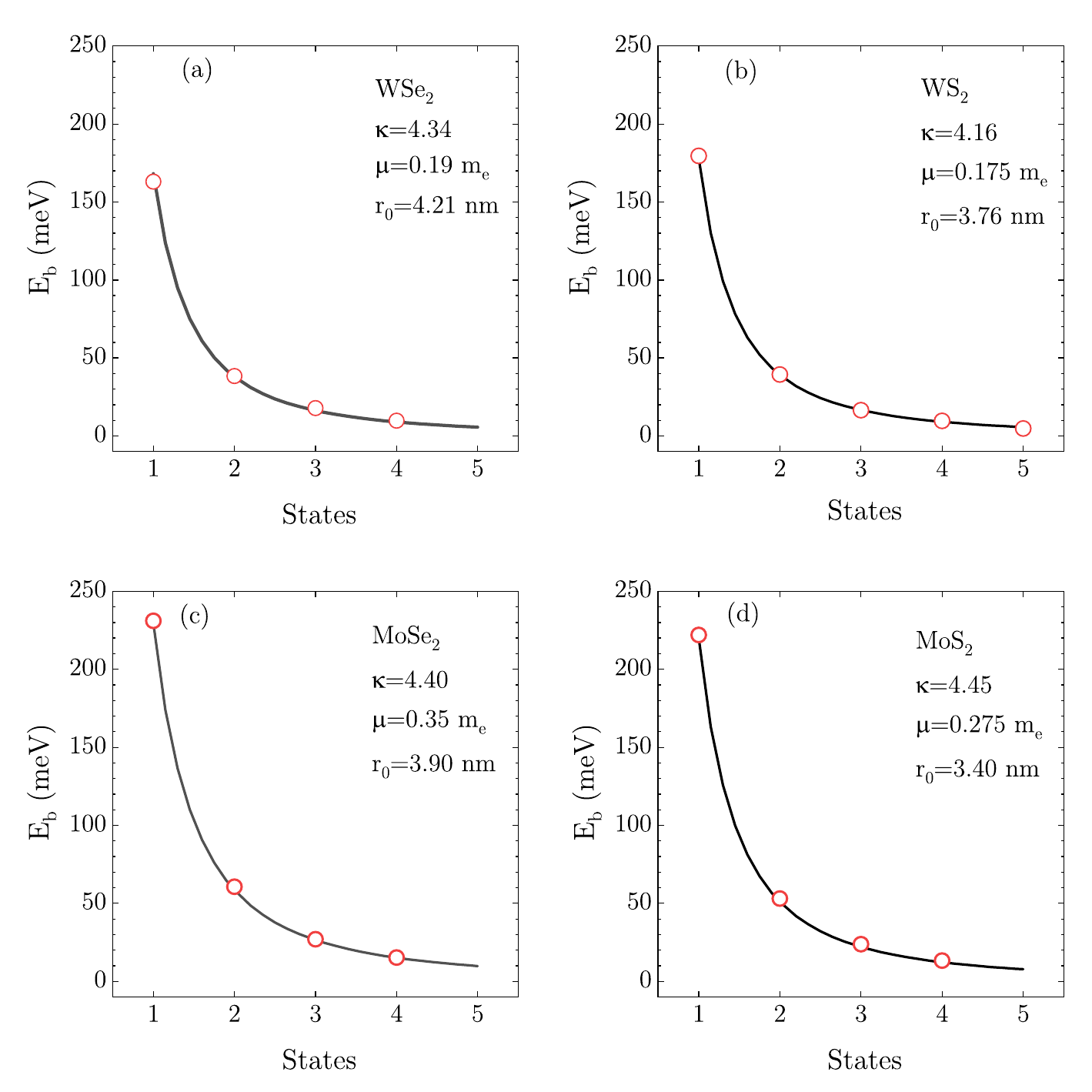}
\caption{Exciton energies for the $ns$ states in monolayers WSe$_2$, WS$_2$, MoSe$_2$, and MoS$_2$ calculated by formula \eqref{s107} (black line) compared with the experimental data \cite{NAT2019, PRL2018} (red circle). Material parameters used for calculations are provided in Table  \ref{tabS1}. }
\label{fig2}
\end{center}
\end{figure} 

\begin{figure}[htbp]
\begin{center}
\includegraphics[width=0.35 \columnwidth]{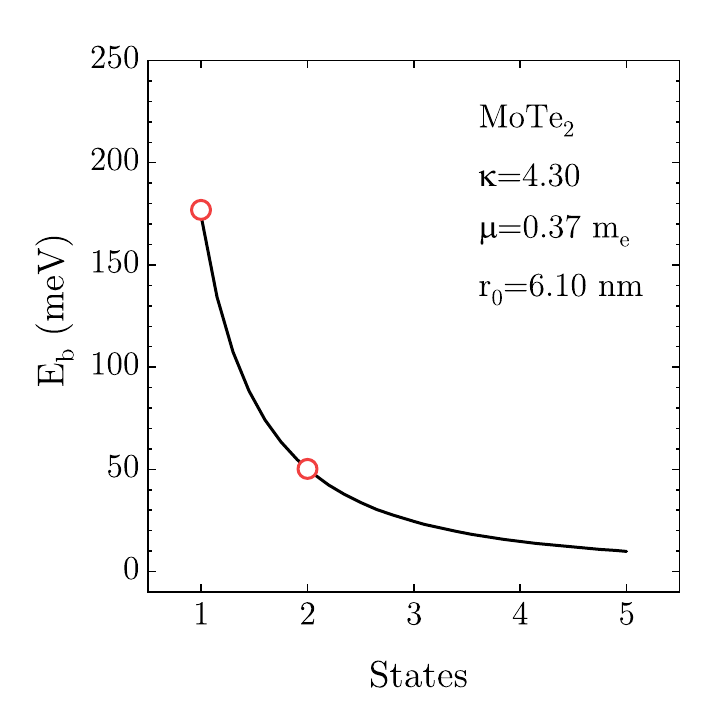}
\caption{Exciton energies for the $ns$ states in monolayer MoTe$_2$ calculated by formula \eqref{s107} (black line) compared with the experimental data \cite{NAT2019} (red circle). Material parameters used for calculations are from the best fit. }
\label{fig3}
\end{center}
\end{figure}

\bibliography{supref}


\end{document}